\documentclass[12pt,preprint,aps,showpacs,floatfix]{revtex4}
\usepackage{epsfig,colordvi}
\usepackage{color}
\usepackage{here}

 \def\as{$\alpha_S\ $}
 \def\T{\textstyle}
 \def\l{\left}
 \def\r{\right}

 \def\be{\begin{equation}}
 \def\ee{\end{equation}}
 \def\bea{\begin{eqnarray}}
 \def\eea{\end{eqnarray}}
 \def\bean{\begin{eqnarray*}}
 \def\eean{\end{eqnarray*}}
 \def\gsim{\mathrel{\rlap{\lower0.2em\hbox{$\sim$}}\raise0.2em\hbox{$>$}}}
 \def\ksim{\mathrel{\rlap{\lower0.2em\hbox{$\sim$}}\raise0.2em\hbox{$<$}}}
 \def\kg{\mathrel{\rlap{\lower0.25em\hbox{$>$}}\raise0.25em\hbox{$<$}}}

\begin{document}

\title{Tomography of a quark gluon plasma at RHIC and LHC energies}

\author{P.B. Gossiaux, R. Bierkandt and J. Aichelin}
\affiliation{SUBATECH, Universit\'e de Nantes, EMN, IN2P3/CNRS
\\ 4 rue Alfred Kastler, 44307 Nantes cedex 3, France}

\date{\today}

\begin{abstract} \noindent
Using the recently published model \cite{Gossiaux:2008jv} for the
collisional energy loss of heavy quarks (Q) in a Quark Gluon Plasma
(QGP), based on perturbative QCD (pQCD) with a running coupling
constant, we study the interaction between heavy quarks and plasma
particles in detail. We discuss correlations between the
simultaneously produced $c$ and $\bar{c}$ quarks, study how central
collisions can be experimentally selected, predict observable
correlations and extend our model to the energy domain of the large
hadron collider (LHC). We finally compare the predictions of our
model with that of other approaches like AdS/CFT.
\end{abstract}


\maketitle

\section{Introduction}
High transverse momentum ($p_T$) single non-photonic electrons which
have been measured in the RHIC heavy ion experiments
\cite{Abelev:2006db,Adare:2006nq} come dominantly from heavy meson
decay. The weighted ratio of their $p_T$ spectra in pp and AA
collisions, $R_{AA}=d\sigma_{AA}/(N_c\
dp_T^2)/(d\sigma_{pp}/dp_T^2)$, where $N_c$ is the number of initial
binary collisions, reveals the energy loss of heavy quarks in the
environment created by AA collisions. Initially the azimuthal
distribution $d\sigma/d\phi \propto (1+2 v_1\cdot \cos(\phi) + 2
v_2\cdot \cos(2\phi))$ of light quarks and gluons is isotropic and
the anisotropy develops during the expansion as an image of the
initial eccentricity in coordinate space. The heavy quarks - created
in a hard process - are initially isotropically distributed in the
transverse momentum space. The final $v_2$ of heavy quarks shows
therefore how the anisotropy of the light quarks and gluons is
transferred to heavy quarks and hence reflects this interaction at
later times.

Recently we have published an approach \cite{Gossiaux:2008jv} in
which we have shown that the energy loss as well as the $v_2 (p_T)$
distribution of the single non-photonic electrons in heavy ion
reactions can be understood in a pQCD based model in which the heavy
quarks interact with the expanding quark gluon plasma (QGP). In
contradistinction to former approaches this model has two
improvements, 1) a running coupling constant and 2) an infrared
regulator in the t-channel, which has been determined by physical
requirements.

It is the purpose of this article to explore the details of the
interaction between the heavy quarks and the QGP in this model, to
determine the consequences for observables, to explore whether there
is a simple way to describe the energy loss in this complicated
environment and to predict correlations between the simultaneously
produced $c$ and $\bar{c}$ quarks. Furthermore we extend the model
to the future LHC collider energies and confront the results with
other theories like the AdS/CFT approach.

\section{The model}
The model \cite{Gossiaux:2008jv} to describe the momentum
distribution of heavy quarks or heavy mesons produced in
ultrarelativistic heavy ion collisions has five major parts which we
will describe one after the other: 1) the initial distributions of
the heavy quarks, 2) the description of the expanding quark gluon
plasma, 3) the elementary interaction between the heavy quarks and
light quarks or gluons, 4) the interaction of the heavy quarks with
the plasma and 5) the hadronization of heavy quarks into open charm
and open beauty mesons.

\subsection{Initial distribution of the heavy quarks}
For the momentum space distribution as well as for the relative
contribution of charm and bottom quarks in pp collisions we use the
pQCD results in fixed order + next to leading logarithm (FONLL) of
Cacciari et al.\cite{cac}.

\begin{figure}[htb]
\epsfig{file=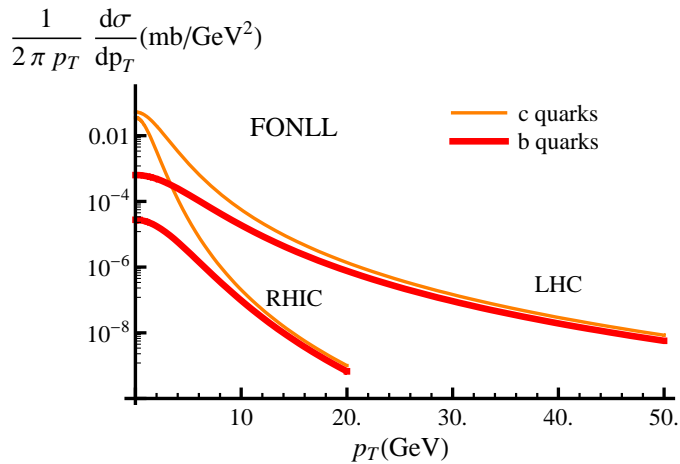,width=0.55\textwidth} \caption{(Color online)
Transverse momentum distribution of c- and b-quarks in fixed order +
next to leading logarithm (FONLL) for RHIC (\cite{Cacciari:2005rk})
and LHC (\cite{cacpri}). } \label{pts}
\end{figure}
For RHIC energies these results have been published in
\cite{Cacciari:2005rk}. They predict a ratio of
$\sigma_{\bar{b}b}/\sigma_{\bar{c}c} = 7 \ 10^{-3}$. Nevertheless,
above $p_T > p_{T\,{\rm cross}} \approx 4\, {\rm GeV}$ more
electrons are produced by B-meson decay than by D-meson decay. The
uncertainty of $p_{T\,{\rm cross}}$ is, however, considerable. This
is due to the uncertainty of the quark masses and of the
factorization and renormalization scales, $\mu_F$ and $\mu_R$. In
this work we have taken $m_c$=1.5 GeV, $m_b=5.1$ GeV and retained
the values of $\mu_R$ and $\mu_F$ that corresponds to the top curves
in the uncertainty bands \cite{Cacciari:2005rk} and that are shown
in Fig. \ref{pts}. They provide, after fragmentation into D and B
mesons and subsequent semi-leptonic decay, the closest agreement
with RHIC non-photonic single electron data. For LHC energies the
initial spectra is stiffer \cite{cacpri} and as well shown in
Fig.\ref{pts}. The heavy quarks are isotropically distributed in
azimuthal direction and therefore their $v_2$ is initially zero. Any
observed anisotropy of heavy mesons is due to the interaction of
their constituents with the medium, and can therefore be used to
reveal the strength of this interaction.

In the E866 experiment at Fermi Lab \cite{E866} it has been observed
that in pA collisions $J/\psi$ mesons have a larger average
transverse momentum as compared to pp collisions. This effect,
called Cronin effect, can be parameterized as an increase of
$<p_T^2>$ by $\delta_0\approx 0.2\ {\rm GeV}^2$ per collision of the
incident nucleon with one of the target nucleons. We have the option
to include this effect by convoluting the initial
transverse-momentum distribution of the heavy quark
\cite{Cacciari:2005rk} with a gaussian distribution of r.m.s
$\sqrt{n_{\rm coll}(\vec{r}_{\perp})\,\delta_0}$. In this
parametrization $n_{\rm coll}$ is taken as the mean number of soft
collisions which the incoming nucleons have suffered prior to the
formation of the $Q\bar{Q}$ pair at transverse position
$\vec{r}_{\perp}$. It turned out that the Cronin effect influences
the $p_T$ spectra below $p_T \approx  5~{\rm GeV}$ but is without
importance for higher $p_T$.

In coordinate space the initial distribution of the heavy quarks is
given by a Glauber calculation.

\subsection{The expanding plasma} The expanding plasma is described
by a hydrodynamical approach neglecting an eventually existing hard
component created by jets. We use the boost invariant model of Heinz
and Kolb which has been described in detail in \cite{heko}. This
model reproduces a variety of experimental findings. Corresponding
to two different equations of state this approach allows to
calculate two distinct scenarios of the expansion. Either the
transition from the QGP to the hadron phase is sudden or the system
traverses a mixed phase. Hadronization after the mixed phase
reproduces the spectra of light mesons and is therefore favored by
experimental data. Without a mixed phase also for heavy quarks  the
interaction time is too short \cite{Gossiaux:2008jv} in order to
reproduce the energy loss and the azimuthal anisotropy seen in the
experimental RHIC data.

Therefore, we use here the model in the mixed phase scenario. We
parameterize the temperature $T(r,t)$ and the mean velocity $u(r,t)$
of this calculation. This quantities serve then to calculate the
interaction of the heavy quarks with the medium. They allow to
calculate the number density of the plasma particles (and hence of
the collision rate) as well as their momentum distribution.

At RHIC the initial entropy density for the hydrodynamical
calculations is chosen in that way that the experimental
multiplicity $dN_{ch}/dy(y=0)$ is reproduced \cite{heko}. For the
LHC prediction we assume that the soft (thermalized) component
contains $1600 <dN_{\rm ch}/dy(y=0) < 2200$. \footnote{What
corresponds, according to the hydrodynamical calculation, to a
plasma lifetime of $6.6 {\rm fm/c}<\tau_{QGP}< 7.4 {\rm fm/c}$ and
to an additional lifetime of the mixed-phase of $\approx$ 4 fm/c.}

\subsection{
Elementary interaction between the heavy quarks and the plasma
particles} Using a fixed coupling constant and the Debye mass
($m_D\approx g_ST$) as infrared regulator pQCD calculations are not
able to reproduce the data, neither the energy loss nor the
azimuthal distribution characterized by $(v_2)$. The novelty of the
approach of ref \cite{Gossiaux:2008jv} is a new description of the
interaction between the heavy quarks and the plasma. As compared to
former pQCD calculations we have introduced a) An effective running
coupling constant, $\alpha_{\rm eff}(Q^2)$, determined from electron
positron annihilation and non leptonic decay of $\tau$ leptons. b)
An infrared regulator in the t channel which is adjusted to give the
same energy loss as calculated in a hard thermal loop approach. In
standard pQCD calculations \cite{Svetitsky:1987gq,Combridge:1978kx}
the gluon propagator in the t-channel Born matrix element has to be
IR regulated by a screening mass $\mu$ \be
    \frac\alpha t \to \frac{\alpha}{t-\mu^2} \, .
\label{aborn} \ee Frequently the IR regulator is taken as
proportional to the square of the Debye mass, $m_D$, \cite{pesh}
 \be \mu^2= {m_D^2}
    = \frac{N_c}{3} \l( 1+\T\frac16\, n_f \r) 4 \pi \, \alpha_S \, T^2\approx(g_ST)^2\,
    \label{eq: mg}\ee
where $n_f\, (N_c)$ are the number of flavors (colors), $g_S^2
=4\pi$\as and T being the temperature. Other approaches use the
square of the thermal gluon mass, $\frac {m_D^2}{3}$ \cite{brad}. In
short, $\mu^2$ is not well determined. Braaten and Thoma
\cite{Braaten:1991jj} have shown for QED that in a medium with
finite temperature the Born approximation is not appropriate for low
momentum transfer $|t|$ but has to be replaced by a hard thermal
loop calculation. Extending their work to QCD we have shown
\cite{Gossiaux:2008jv} that the energy loss, calculated with pQCD
matrix elements of the form of eq. \ref{aborn}, only agrees with
that calculated in a hard thermal loop approach if $\mu^2$ is much
smaller and around \be \mu^2\approx 0.2 g_S^2\,T^2. \label{IR}\ee
Employing a running coupling constant and replacing the Debye mass
by an effective IR cut off (eq. \ref{IR}) we find a substantial
increase of the collisional energy loss which brings for the RHIC
experiments $v_2(p_T)$  as well as $R_{AA}(p_T)$ to values close to
the experimental ones without excluding a contribution from
radiative energy loss. More precisely, the collisional cross section
has to be multiplied by a K-factor of around 2 (which is assumed to
be identical for c- and b-quarks) in order to reproduce the data.
Thus the difference to the data is of the order that we expect for
the contribution from radiation energy loss which is not included
here.

In this article, we follow the labeling established in ref.
\cite{Gossiaux:2008jv}. The approach with a running coupling
constant is dubbed ``model E''. In order to point out the influence
of the running coupling constant we present also some results for
the so-called ``model C'', in which the coupling constant is taken
as $\alpha_s(2\pi T)$ and $\mu^2 = 0.15 m_D^2$. This model requires
$K\approx 5$ to reproduce the RHIC data. In all calculations,
presented here, the corresponding K-factors have been applied.

\subsection{Interaction of the heavy quark with the expanding
plasma} The heavy quarks can scatter elastically with the gluons and
light quarks which are present in the QGP. The temperature field,
determined by the hydrodynamical calculations, allows to calculate
the density and - together with the local expansion velocity of the
plasma - the momentum distribution of the light quarks and gluons
which scatter with the heavy quark. The interaction is described by
a Boltzmann equation which is solved by the test particle method,
applying Monte Carlo techniques. For the collisions between the
heavy quarks and the plasma particles we apply the elementary pQCD
cross sections. We follow the trajectory of the individual heavy
quarks from creation until hadronization but do not pursue that of
the plasma particles. The hadronization happens when the energy
density of the fluid cell falls under a critical value of the energy
density. This is $1.64~{\rm GeV}/{\rm fm}^3$ in the scenario without
mixed phase and $0.5~{\rm GeV}/{\rm fm}^3$ at the end of the mixed
phase. It is assumed that after hadronization heavy mesons do not
interact with the hadronic environment.

\subsection{Hadronization}
\begin{figure}[H]
\begin{center}
\includegraphics[width=8cm]{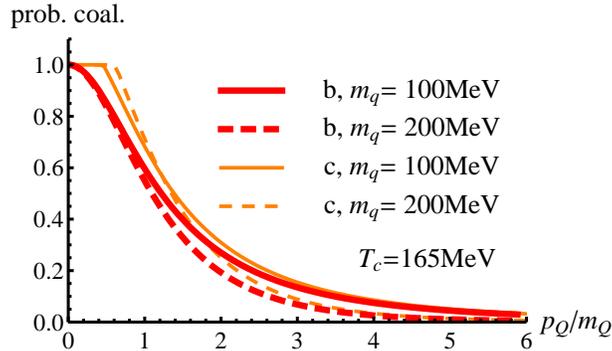}
\end{center}
\caption{(Color online) Relative contribution of coalescence of a
c(b) quark with a light quark at freeze out to the D(B) meson yield
as a function of the relative momentum $p_Q/m_Q$ of the heavy quark.
Heavy mesons which are not produced by coalescence are created by
fragmentation as described in \cite{Cacciari:2005rk}.}
\label{graph1}
\end{figure}

The heavy quarks form hadrons either by coalescence or by
fragmentation. In our calculation the relative fraction depends on
$p_Q$, on the fluid cell velocity and on the orientation of the
hadronization hypersurface $\Sigma$ as explained below. The
coalescence mechanism is based on the model of Dover
\cite{Dover:1991zn}. To describe the creation of a heavy meson by
coalescence we start from
\begin{eqnarray}
N_{\Phi={D,B}}&=&\int p_Q\cdot d\sigma_1\, p_q\cdot d\sigma_2
\frac{d^3p_Q}{(2\pi \hbar)^3 E_Q}\frac{d^3p_q}{(2\pi \hbar)^3 E_q}
\nonumber\\
&&\hspace{1cm}\times f_Q(x_Q,p_Q)f_q(x_q,p_q)
f_\Phi(x_Q,x_q;p_Q,p_q)
\end{eqnarray}
where $f_Q$ et $f_q$ are normalized to the number of quarks which
go through the hypersurface:
\begin{equation}
\int p_Q\cdot d\sigma_1\, \times f_Q(x_Q,p_Q) \frac{d^3p_Q}{(2\pi
\hbar)^3 E_Q}=N_Q=1
\end{equation}
for hadronization of a given heavy quark and
\begin{equation}
\int p_q\cdot d\sigma_2\, \times f_q(x_q,p_q) \frac{d^3p_q}{(2\pi
\hbar)^3 E_q} =N_q.
\end{equation}
$f_\Phi$ is the invariant probability density that a heavy quark at
the position $x_Q$ with momentum $p_Q$ forms a heavy meson $\Phi$
with a light quark with $x_q, p_q$, which traverses the hypersurface
$\Sigma$. $f_q(x_q,p_q)$ is the distribution of the light quarks
which is assumed to be a thermal Boltzmann-J\"uttner distribution.
Assuming that $f_\Phi$ factorizes we use
\begin{equation}
f_\Phi(x_Q,x_q;p_Q,p_q)=\exp\left(\frac{(x_q-x_Q)^2-((x_q-x_Q)\cdot
u_Q)^2} {2 R_c^2}\right)\times F_\Phi(p_Q,p_q).
\end{equation}
$u_Q$ is the 4-velocity of the heavy quark. Thus in the rest system
of the heavy quark $f_\Phi$ is a Gaussian function of
$\|\vec{x}_q-\vec{x}_Q\|$. Coalescence requires that in coordinate
space the position of the heavy and of the light quark are very
close and we obtain
\begin{equation}
N_{\Phi}=\int \frac{d^3p_q}{(2\pi \hbar)^3 E_q} \frac{p_q\cdot
\hat{d\sigma}}{u_Q \cdot \hat{d\sigma}} f_q(x_Q,p_q) (\sqrt{2\pi}
R_c)^3F_\Phi(p_Q,p_q)\,,
\end{equation}
where $\hat{d\sigma}$ is the unit vector along $d\sigma$. The
$u_Q\cdot \hat{d\sigma}$ denominator is positive if the heavy quark
escapes from the plasma. It counts for the fact that a heavy quark
coming out tangentially to the critical hypersurface $\Sigma$ has a
larger chance to encounter its light partner. For $F_\Phi(p_Q,p_q)$
we take
\begin{equation}
F_\Phi(p_Q,p_q)= \exp \left(\frac{(\frac{p_Q}{m_Q} -
\frac{p_q}{m_q})^2} {2\alpha_d^2}\right).
\end{equation}
The coalescence probability is maximal if $x_q=x_Q$ and  $p_q=p_Q$.
The normalization condition   \begin{equation} \int \frac{d^3 r d^3
p}{(2\pi \hbar)^3}f_\Phi(x_Q,x_q,p_Q,p_q)=1,\end{equation} where $r$
and $p$ are the relative coordinates between Q and q, relates
$\alpha_d^2$ and $R_c^2$. We find
\begin{equation}
N_{\Phi}(x_Q;p_Q) = \frac{\tilde{c}_d g_q}{u_Q\cdot
\hat{d\sigma}(x_Q)} \int_{u_q\cdot\hat{d\sigma}(x_Q)>0} \frac{d^3
u_q}{u_0}\, u_q\cdot \hat{d\sigma}(x_Q)\,
e^{-\left(\frac{m_q}{T_c}\,u_{\rm cell}(x_Q)+
\frac{u_Q}{\alpha_d^2}\right)\cdot u_q}\,, \label{dover_nphi}
\label{coalprob}
\end{equation}
where $g_q$ is the degeneracy factor of the light quarks, $u_q$ is
their 4-velocity and
\begin{equation}
\tilde{c}_d:= \left(\frac{m_Q+m_q}{m_Q}\right)^3\times \frac{1}{4\pi
\alpha_d^2 K_2\left(\frac{1}{\alpha_d^2}\right)} \approx
\frac{1}{4\pi \alpha_d^2 K_2\left(\frac{1}{\alpha_d^2}\right)}
\end{equation}
if $m_Q\gg m_q$. For the calculation we assume a critical
temperature of $T_c=165~{\rm MeV}$. Equation \ref{coalprob} is up to
a factor the Cooper-Frye formula which describes the hadronization
of particles at the surface of the expanding plasma, with an
effective inverse temperature $\beta_{\rm eff}$ and an effective
4-velocity $u_{\rm{cell,eff}}$ such that $\beta_{\rm eff}
u_{\rm{cell,eff}} = \beta_c u_{\rm{cell}}+ \frac{u_Q}{\beta_c m_q
\alpha_d^2}$. For a given choice of the mass $m_q$, we complete our
coalescence algorithm by fixing $\alpha_d$ in such a way
\footnote{This leads to $\alpha_d=0.88$ for $m_q=100$ MeV and to
$\alpha_d=0.39$ for $m_q=200$ MeV.} that $N_B=1$ for a b-quark at
rest in a fluid cell with $\hat{d\sigma}=u_{\rm cell}$, in agreement
with the physical picture that such a heavy quark can hadronize
exclusively by coalescence. The numbers $N_D$ and $N_B$ calibrated
in this way are thus interpreted as coalescence {\em probabilities},
illustrated in Fig. \ref{graph1}.
For momenta above $p_Q = 0.5~{\rm GeV}$ the probability to form a
heavy meson by coalescence falls below one. Because all heavy quarks
appear finally as heavy mesons we assume that all heavy quarks which
do not coalesce form  mesons by fragmentation, as described in
\cite{Cacciari:2005rk}. As one can see in Fig. \ref{graph1}, high
$p_T$ heavy mesons are formed exclusively by this mechanism.

\begin{figure}[H]
\begin{center}
\epsfig{file=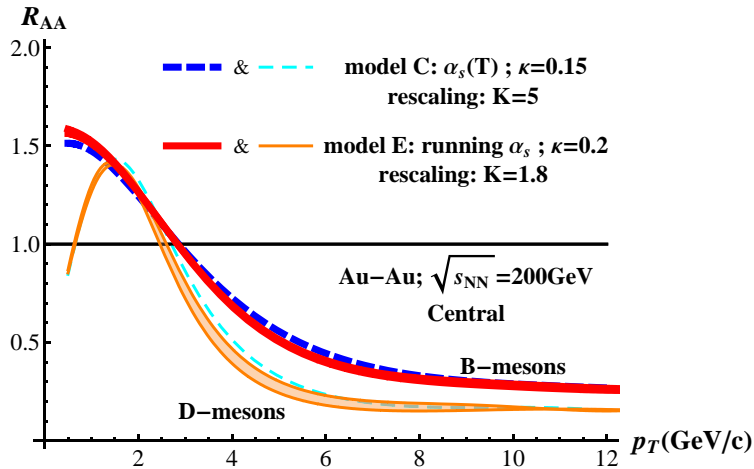,width=0.6\textwidth}
\end{center}
\caption{(Color online) $R_{AA}$ as a function of $p_T$ for D- and
B- mesons. We display  $R_{AA}$ for two parameterizations which
describe the experimental data after the results have been
multiplied with appropriate K-factors (see ref.
\cite{Gossiaux:2008jv} and section II.C for details).} \label{RHIC0}
\end{figure}

By this hadronization procedure we get a good description of the
$p_T$ spectrum over the whole $p_T$ range. The physics can best be
discussed in terms of $R_{AA}$ which is expected to be one if no
medium is present. Our results for the D and B mesons for central
Au+Au collisions at RHIC are displayed in Fig. \ref{RHIC0} (for the
explication of the different models we refer to
\cite{Gossiaux:2008jv}). In this figure, the upper (lower) limit of
the ``D-meson'' band for model E corresponds to $m_q=100$ MeV
($m_q=200$ MeV) in equation \ref{coalprob}. For B-mesons the
difference between the two corresponding curves is of the order of
the line width. In the following, we will retain $m_q=100$ MeV.

\section{Tomography at RHIC energies}
\subsection{Momentum loss}
We start our analysis with the investigation of the momentum loss
\begin{figure}[htb]
\epsfig{file=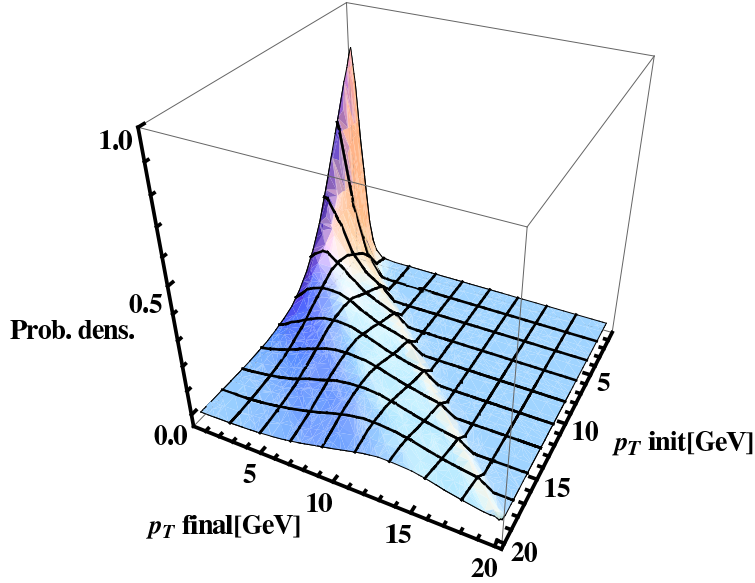,width=0.6\textwidth} \caption{(Color
online) Mean value and variance of the final $p_T$ momentum
distribution of c-quarks as a function of their momentum at
production for central Au+Au collisions at $\sqrt{s} = 200~{\rm
GeV}$.} \label{RHIC1}
\end{figure} the
heavy quarks suffer while traversing the plasma. In Fig. \ref{RHIC1}
we display,  for central Au+Au collisions at $\sqrt{s} = 200~A{\rm
GeV}$, the conditional probability density of the transverse
momentum loss as a function of the initial momentum of the heavy
quarks. At high initial momenta ($p_T > 5~{\rm GeV}$) we observe a
quite broad distribution which narrows for smaller $p_T$ values. For
very low initial momenta we see an increase of the transverse
momentum during the expansion. If their initial $p_T$ value is
smaller than that expected for heavy quarks in equilibrium with
their environment the interactions with the plasma particles
increase their momenta.
\begin{figure}[htb]
\epsfig{file=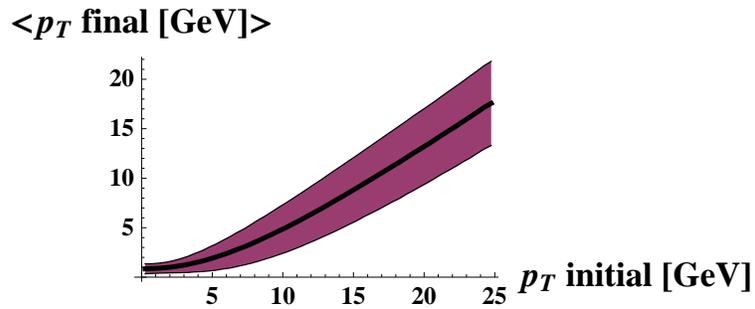,width=0.6\textwidth} \caption{(Color
online) Mean value and variance of the conditional probability
density for a c-quark with a final transverse momentum $p_T$ as a
function of the initial $p_T$.} \label{RHIC2}
\end{figure}
In Fig. \ref{RHIC2} the mean value and the variance of this
probability density are plotted. Above $p_T = 10~{\rm GeV}$ we
observe, despite of the complex path length distribution, to a very
good approximation a linear dependence of the $p_T$ loss on the
initial $p_T$ momentum which can be described by $<p_T^{\rm final}>
= p_T^{\rm initial} - 0.08 p_T^{\rm initial} - 5~{\rm GeV}$.
Numerically, the constant -5 GeV energy loss dominates the $- 0.08\
p_T^{\rm initial}$ term even for high $p_T^{\rm initial}$ but for
quantitative comparisons the latter is not negligible. This is
consistent with the underlying microscopic energy loss, as
$\frac{dE}{dx}$ was shown to saturate at large initial momenta due
to asymptotic freedom \cite{pesh}. Also the variance depends
linearly on the initial $p_T$ value for high initial momenta. For
low initial momenta the situation becomes more complex, the final
$p_T$ approaches there the value expected for heavy quarks in
equilibrium with their environment.

\begin{figure}[H]
\begin{center}
\includegraphics[width=7cm]{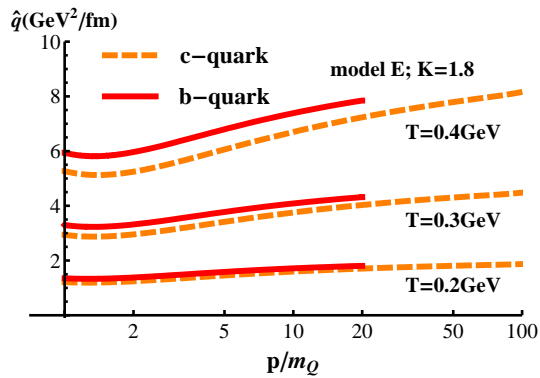}
\end{center}
\caption{(Color online) $\hat q$ as a function of the momentum of
the heavy quark for the standard parameter set E and for three
temperatures of the plasma. \cite{Gossiaux:2008jv}.} \label{qhat}
\end{figure}

To allow for a comparison with other approaches, it is interesting
to make the link between the energy loss of our model and the
transport coefficient \be \hat q = \frac{<k^2_\perp>}{\lambda}=
<k^2_\perp> \sigma \rho= \frac{1}{v_Q}\frac{<k^2_\perp>}{\Delta
t}\approx \frac{4B_\perp}{v_Q} \ee which describes the average
squared transverse momentum transfer in a single collision divided
by the mean free $\lambda. \ \sigma, \rho, B_\perp, \Delta t$ and
$v_Q$ are the heavy quark parton cross section, the parton density
in the medium, the transverse diffusion coefficient
\cite{Gossiaux:2008jv}, the time between two subsequent collisions
and the heavy quark velocity, respectively. Fig. \ref{qhat} shows
$\hat q$ as a function of the momentum of the heavy quark.
\subsection{Dependence of the momentum loss on the creation point in
coordinate space}
The momentum loss of a heavy quark depends on the creation point of
the heavy quark - anti-quark pair.
\begin{figure}[htb]
\epsfig{file=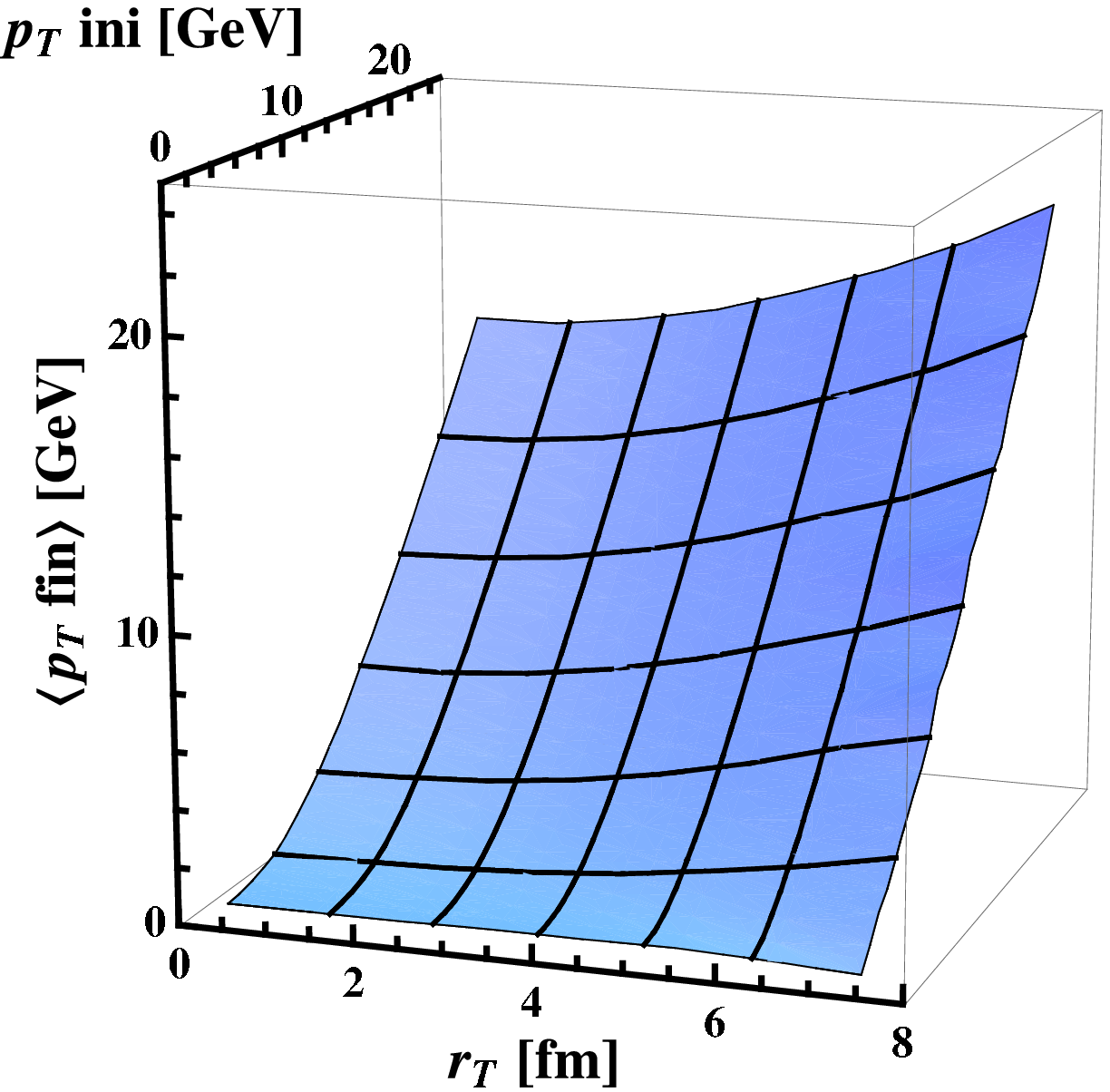,width=0.45\textwidth}
\epsfig{file=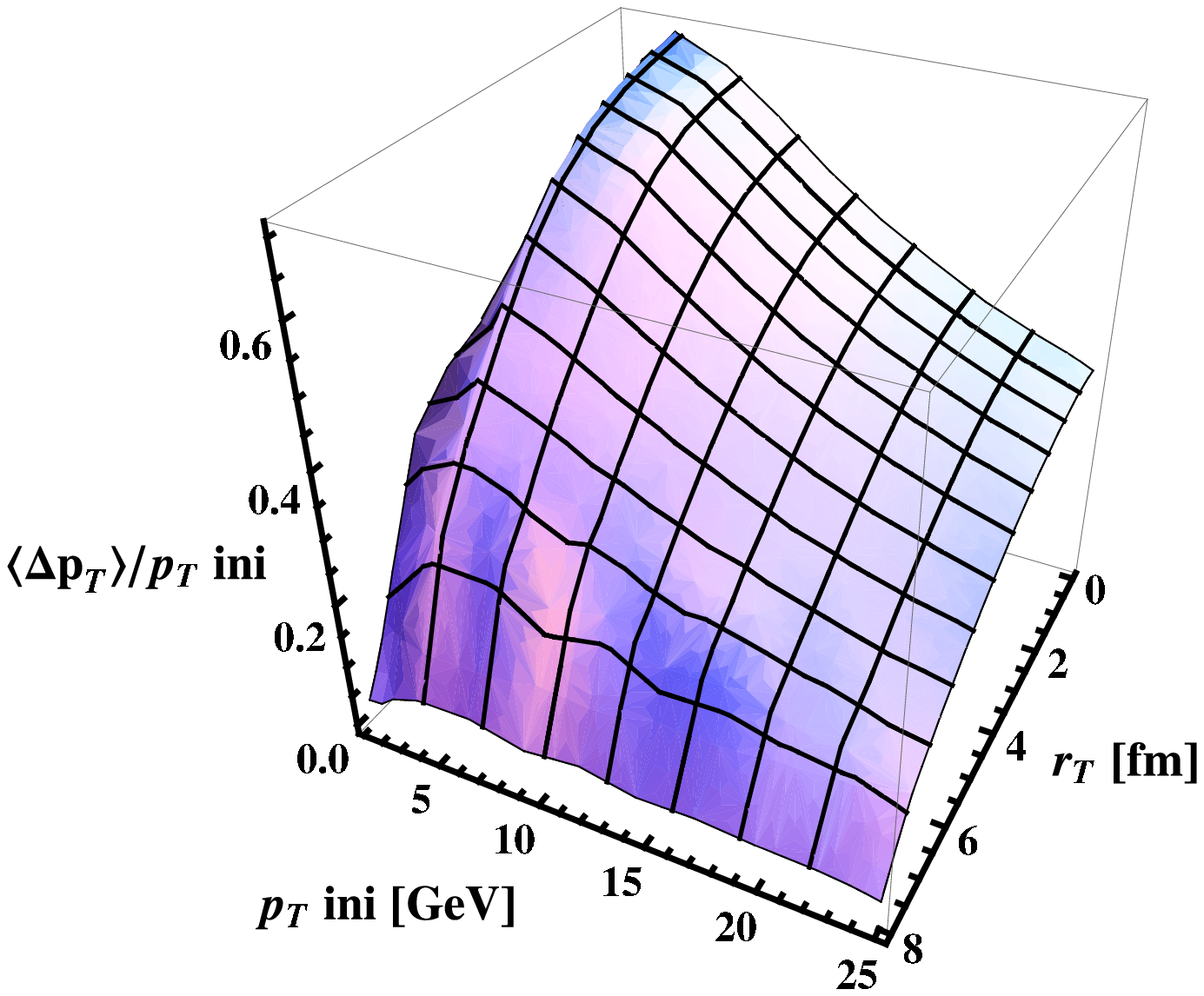,width=0.45\textwidth}
\caption{(Color online) Left: Average final momentum of the c-quark as a
function of its initial momentum and of the centrality of its creation point.
Right: Relative momentum loss of the heavy quark as a function of
its initial momentum and of the centrality of its creation point.
All calculations are done for central Au+Au collisions at $\sqrt{s}
= 200~{\rm GeV}$.}
\label{RHIC3}
\end{figure}
If one wants to know the information about the QGP, which is
contained in the measured $p_T$ spectra, it is important to know
from which part of the QGP the observed heavy quarks originate. Fig.
\ref{RHIC3} shows, on the left hand side, the average final $p_T$ of
the heavy (anti)quarks as a function of their initial $p_T$ and of
the transverse distance of their creation point with respect to the
center of the reaction, $r_T^{in}$. We see three different regimes:
at low initial $p_T$ the average final momentum is independent of
the distance to the center. These are heavy quarks which come to an
equilibrium with the environment. At large values of $r_T^{in}$ the
momentum loss is  small because the heavy quarks are too close to
the (in the hydrodynamical calculation shrinking) surface to
interact really with the plasma. The third type of heavy quarks are
those which have initially a high momentum and have been created
close to the center. These quarks are really penetrating probes
traversing an important fraction of the plasma. The momentum loss of
these particles is high but it does not change substantially between
$0 < r_T^{in}<4 $fm. In other words, Fig. \ref{RHIC3} shows an
explicit path-length dependence, despite the rapid decrease of the
energy density during plasma cooling. This fact contradicts the
conclusion of \cite{Huang:08}. A complimentary view of the
centrality dependence of the momentum loss is shown on the right
hand side of Fig. \ref{RHIC3}. There we plot the relative momentum
loss as a function of the initial transverse momentum $p_T$ and of
the centrality. The relative momentum loss increases with centrality
and with decreasing initial momentum. Heavy quarks with a moderate
initial momentum suffer the heaviest relative momentum loss. There
the kinematics of the collisions allows for large angle scattering
and therefore to a relatively fast approach to equilibrium. The
kinematics of the collisions of fast heavy quarks with the thermal
environment leads only to a moderate momentum transfer and therefore
the direction of the heavy quark changes only little.
\begin{figure}[htb]
\epsfig{file=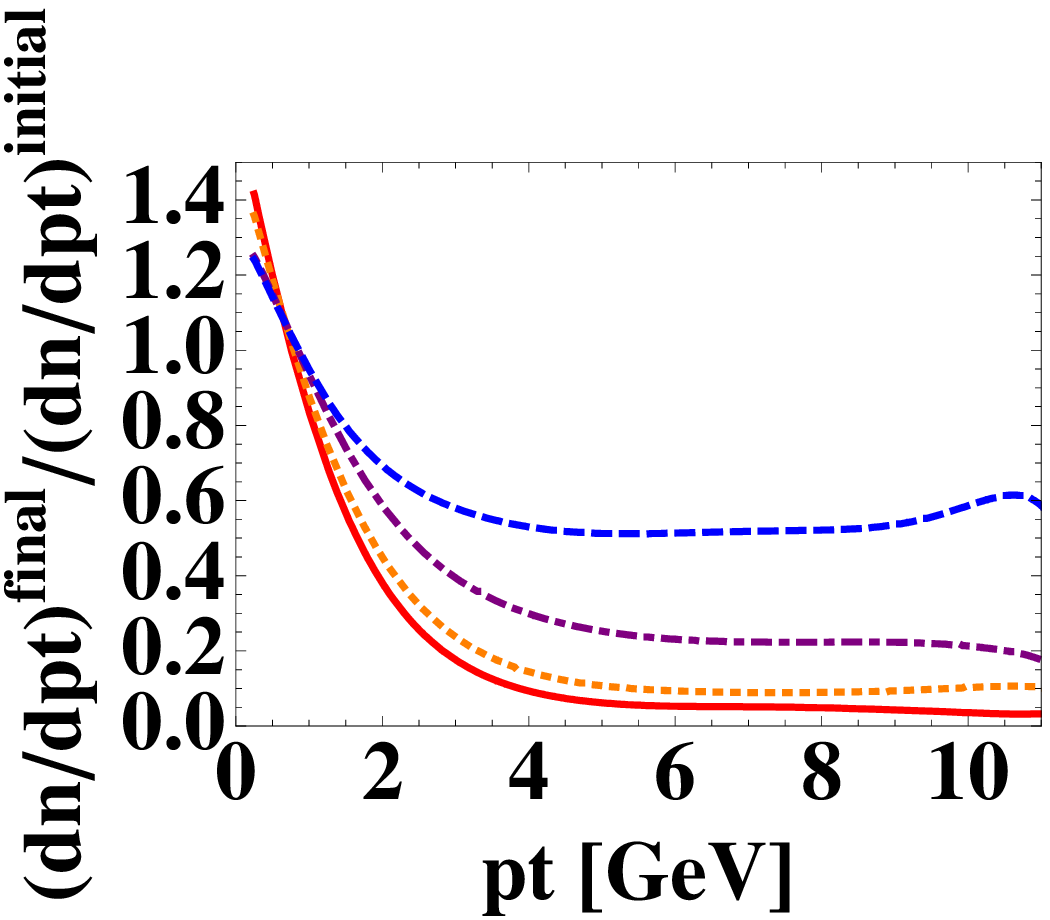,width=0.45\textwidth}
\epsfig{file=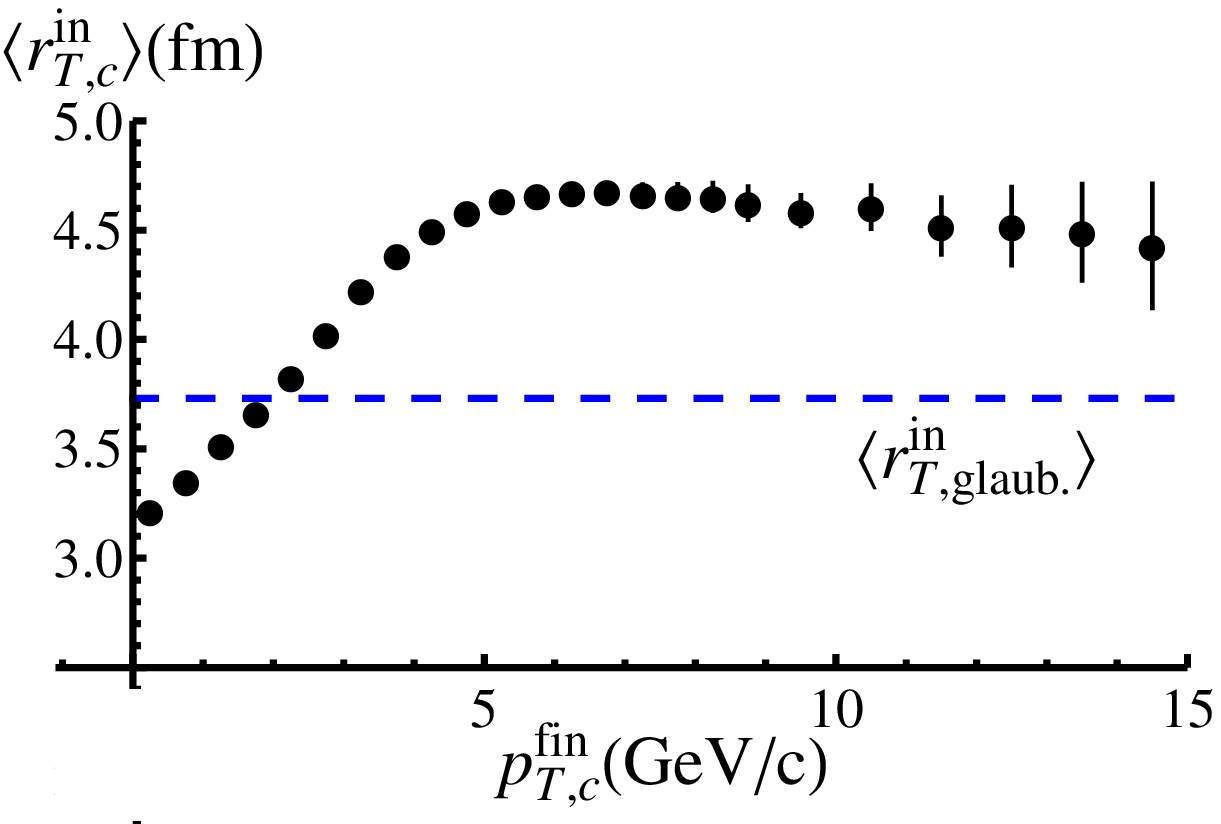,width=0.45\textwidth}
\caption{(Color online) Left: $(dN/dp_T^{\rm fin(al)})/
(dN/dp_T^{\rm in(itial)})$ of c-quarks produced at a [0-2(full),
2-4(short-dashed), 4-6(dashed-dotted), 6-8(dashed)] [fm] transverse
distance to the ``center'' (symmetry axis) of the reaction for
central Au+Au collisions at $\sqrt{s} = 200~{\rm GeV}$. Right:
average transverse position of the production points of the c-quarks
as a function of their final momentum. The dashed line corresponds
to the $p_T$ averaged position. The bars mark the statistical
uncertainties in the simulations.} \label{RHIC4}
\end{figure}
Fig. \ref{RHIC4} (left) shows the dependence of $(dN/dp_T^{\rm
fin(al)}/(dN/dp_T^{\rm in(itial)})$ on the production point of the
charmed quark - anti-quark pair. Pairs produced at the center of the
reaction are highly suppressed at large $p_T$. Therefore quarks
which contribute in this kinematical regime are predominantly from
the surface and contain little information on plasma properties in
the center of the reaction. This corona effect can be illustrated
alternatively using the correlation between the average initial
transverse position and the final transverse momentum
\footnote{While in the dynamical evolution, it is the initial
position that (partly) determines the heavy quark evolution and its
final momentum before hadronization, it is quite natural, in this
type of ``reverse analysis'', to perform selections on final
observables and to investigate how they permit to access former
properties of the distribution.} displayed in Fig. \ref{RHIC4}
(right). There one can see that heavy quarks, passing the
hadronization hypersurface with $p_T \le 3~{\rm GeV}$, originate
from larger initial transverse distances than the overall Glauber
average ($\approx 3.7~{\rm fm}$).  Heavy quarks with a small final
transverse momentum come from production points which are more
central than the average. We observe a (slow) decrease of $\langle
r_{T}^{\rm in}\rangle$ for large values of $p_T$, because with
increasing $p_{T,c}^{\rm fin}$ the matter becomes less and less
opaque (see Fig. \ref{RHIC3}).

We now discuss a possible way to use the $Q\bar{Q}$ pair as a
trigger to probe inner regions of the QGP. Within our model, each
$Q\bar{Q}$ pair is initially created back to back with the same
momentum \footnote{NLO corrections spoil of course this idealized
view; they are expected to be large at LHC for $c\bar{c}$ pairs.}.
For the most central production points, the final momentum
difference is small because the path lengths in the plasma are
almost the same for both quarks. The more peripheral the pair is
produced the more the effective path lengths in the QGP can be
different. Therefore, the smaller the final $p_T$ difference of the
simultaneously produced c and $\bar{c}$ pair the more it is probable
that it has been produced at a small distance from the center. Our
approach thus predicts a strong correlation between the final
transverse-momentum difference of a given $Q\bar{Q}$ quark pair and
its initial position in the transverse plane. This correlation could
possibly be exploited experimentally to discriminate this model
against other approaches where energy loss is not due multiple
independent collisions.
\begin{figure}[htb]
\includegraphics[scale=0.5]{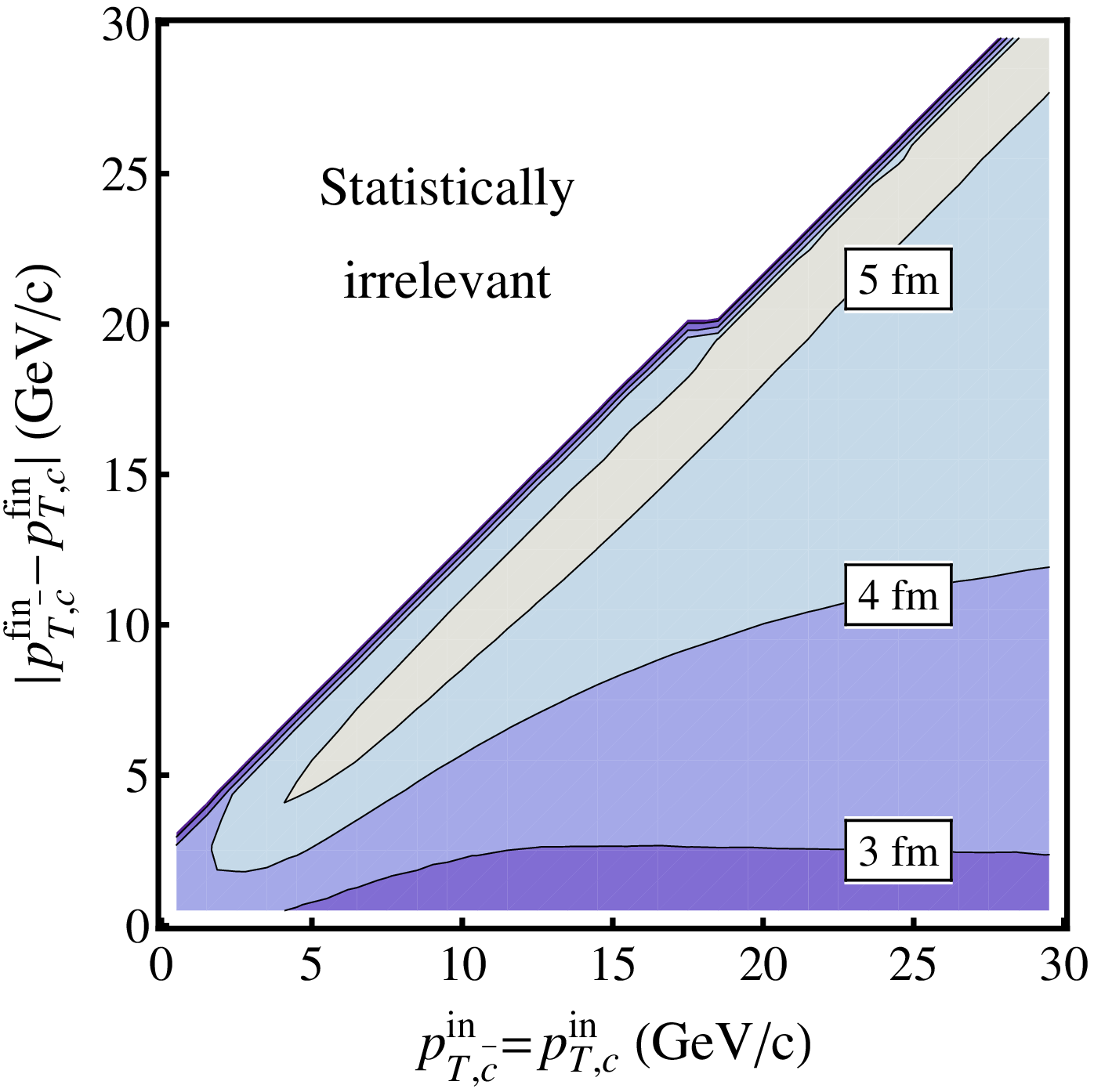}
\includegraphics[scale=0.5]{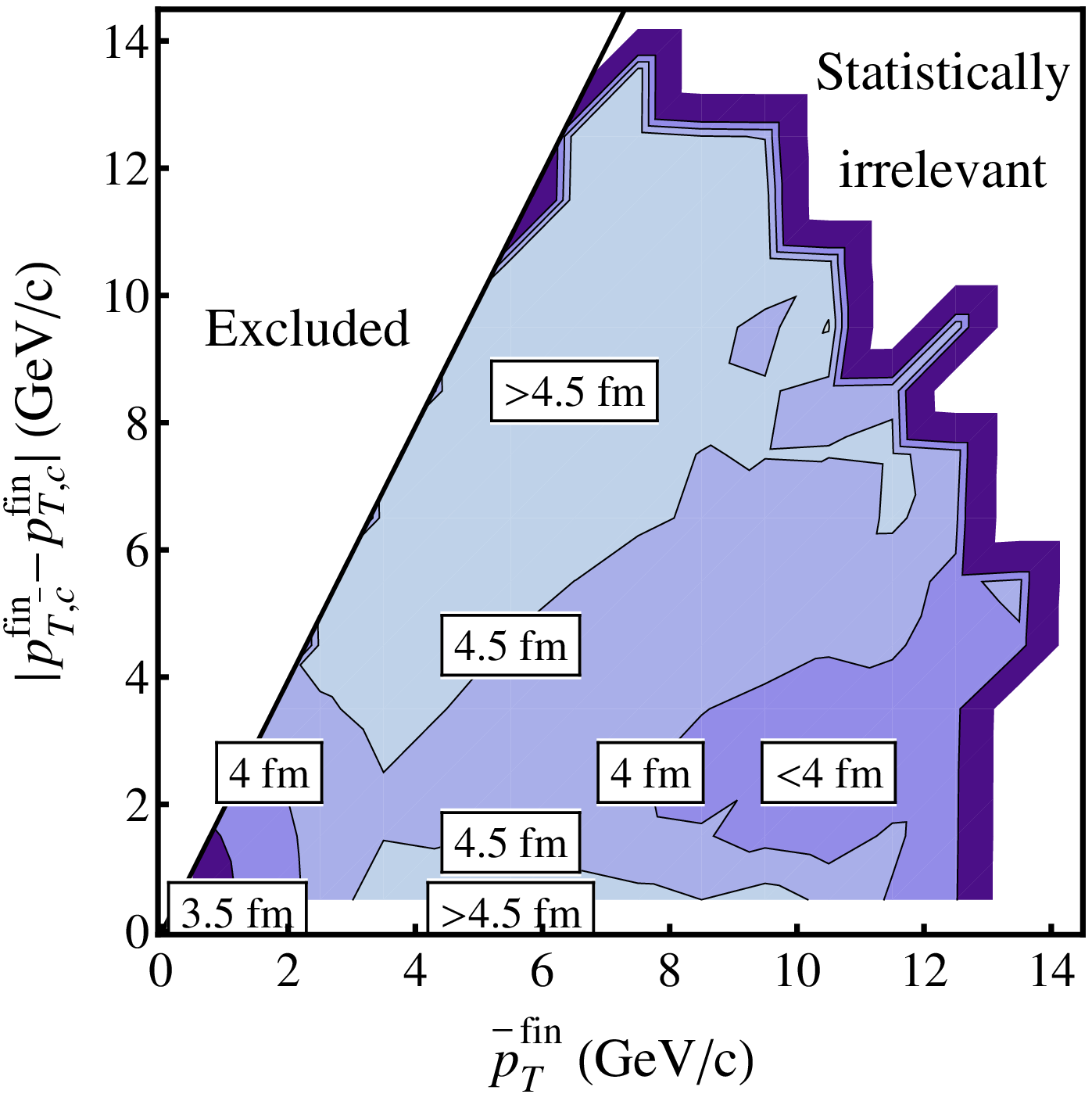}
\caption{(Color online) Left: Correlation between the average
centrality $\langle r_T^{in} \rangle$ of the production points of
the $c\bar{c}$ pair (labeled iso-contours, in [fm]) and the
difference between the final momenta of the quarks $\Delta p_T^{\rm
fin}:=|p_{T,c}^{\rm fin}-p_{T,\bar{c}}^{\rm fin}|$ for various
initial momentum $p_{T,c}^{\rm in}= p_{T,\bar{c}}^{\rm in}$. Right:
same correlation as left for various final pair momenta
$\bar{p}_{T}^{\rm fin}:= \frac{p_{T,c}^{\rm fin}+p_{T,\bar{c}}^{\rm
fin}}{2}$.} \label{RHIC5}
\end{figure}
In Fig. \ref{RHIC5} (left), we study further this correlation for
simultaneously created $c\bar{c}$ pairs. We display the average
transverse distance $\langle r_T^{in} \rangle$  of their production
points to the center of the reaction as a function of the initial
(anti)quark momentum $p_T^{\rm in}$, separated for different values
of the final momentum difference between the $c$ and the ${\bar c}$
quark. For $p_T^{\rm in}> 5\,{\rm GeV}$, one sees the expected
correlation: The quarks of pairs created far from the center
($\langle r_T^{\rm in}\rangle$ large) have usually quite different
path lengths and show therefore finally a large momentum difference
$\Delta p_T^{\rm fin}$. On the contrary, for quarks created close to
the center the path length is similar and therefore $\Delta p_T^{\rm
fin}$ is small. Therefore, by selecting events with small $\Delta
p_T^{\rm fin}$ (w.r.t $p_T^{\rm in}$), one can trigger on more
central events in order to study their properties.

In practice, one does of course not have access to $p_T^{\rm in}$ as
we only measure particles in their asymptotic state. Therefore is is
useful to study the same correlation as a function of the average
between $p_{T,c}^{\rm fin}$ and $p_{T,\bar{c}}^{\rm fin}$, i.e.
$\bar{p}_T^{\rm fin}:=\frac{p_{T,c}^{\rm fin}+p_{T,\bar{c}}^{\rm
fin}}{2}$. This is done in Fig.\ref{RHIC5} (right). Due to the
energy loss the structure has changed as compared to the left panel.
We find now that for intermediate values of $\bar{p}_T^{\rm fin}$
($\in [3,10]$ GeV), requesting $\Delta p_T^{\rm fin}\approx 0$ leads
to {\em larger} values of $\langle r_T^{\rm in} \rangle$, of the
order of 5~fm. A refined analysis shows that this small-$\Delta
p_T^{\rm fin}$ crest is due to pairs which are produced peripherally
and tangentially to the fireball cylinder ($\vec{p}_T^{\rm in} \perp
\vec{x}_T^{\rm in}$). Hence the trajectories of both quarks have
approximately the same path-length in matter. Although these events
are much less frequent than a $c\bar{c}$ production inside the bulk
of the QGP, the associated energy loss is rather small, and
$p_{T,c}^{\rm fin}\approx p_{T,c}^{\rm in}$, so that they dominate
the final spectra due to the steeply falling $\frac{d\sigma_{\rm
prod}}{d p_T^{\rm in}}$.
\begin{figure}[htb]
\includegraphics[scale=0.5]{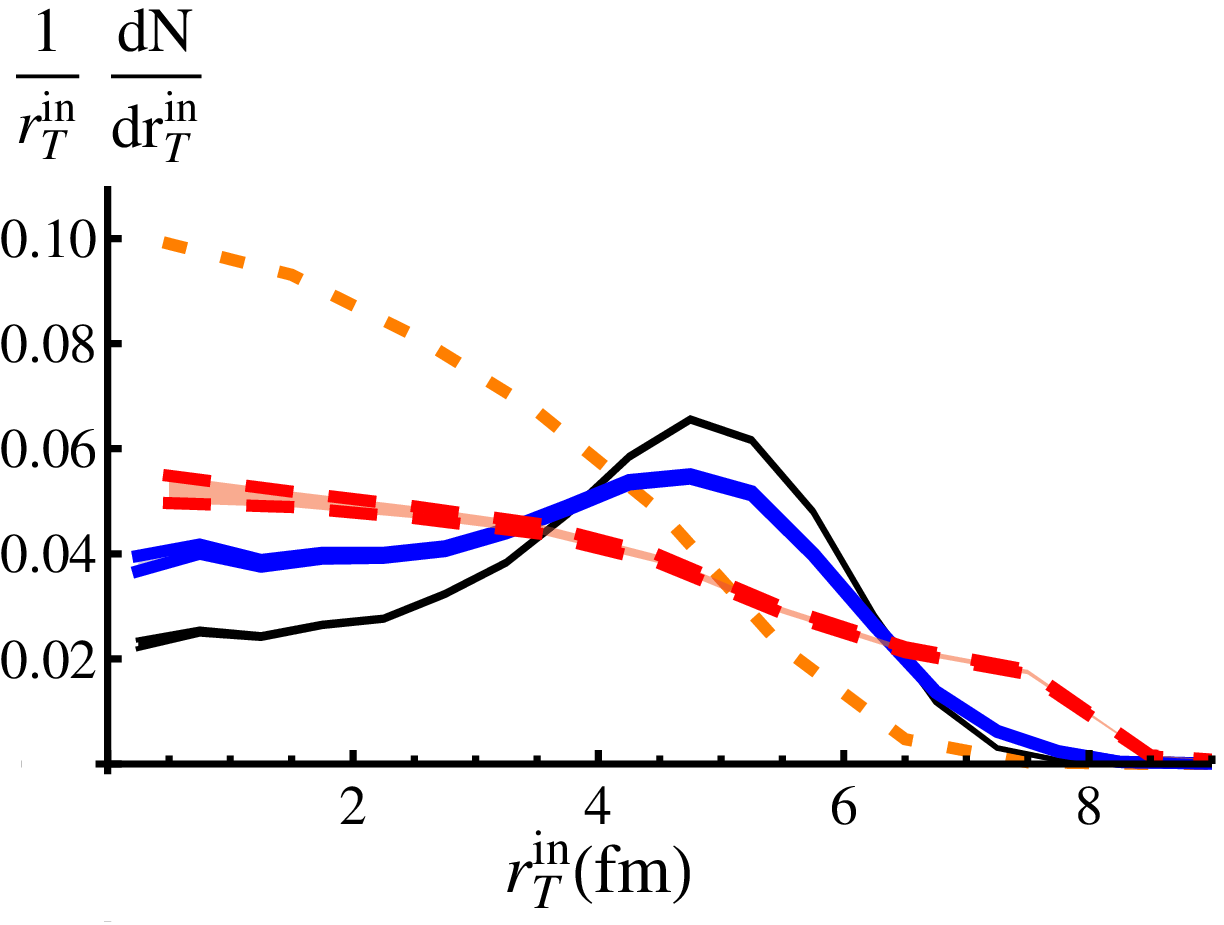}
\includegraphics[scale=0.5]{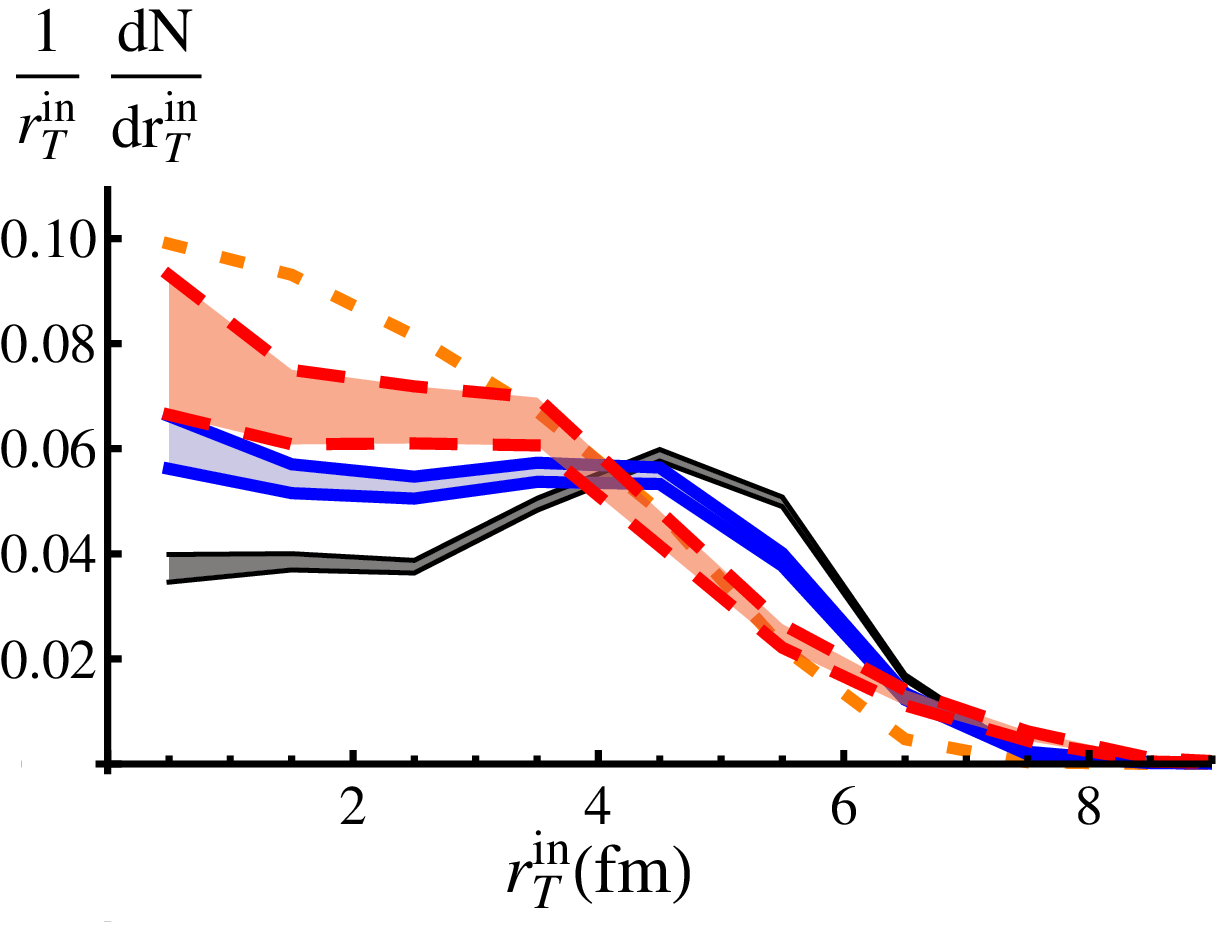}
\caption{(Color online) Left: Distribution of the radial distance
$r_T^{in}$ of the production points of the $c\bar{c}$ pairs for
various conditions on their final momenta: no selection (dotted,
orange), $p_{T,c}^{\rm fin}>5~{\rm GeV}$ (thin, black),
$\bar{p}_{T}^{\rm fin}>5~{\rm GeV}$ (thick, blue) and
$\bar{p}_{T,c}^{\rm fin}>5~{\rm GeV} \cap \Delta p_{T}^{\rm fin}<
0.2\,\bar{p}_{T}^{\rm fin}$ (dashed, red). All distributions are
normalized to unity. Right: same, with a lower bound of $p_T$ of 10
GeV instead of 5 GeV.} \label{RHIC5bis}
\end{figure}
This interpretation is confirmed by analyzing the radial
distribution of the creation points, $r_T^{\rm in}$, for different
conditions on $\bar{p}_{T,c}^{\rm fin}$, as shown in Fig.
\ref{RHIC5bis} (left). For large final $p_T$ values, $p_{T,c}^{\rm
fin}>5~{\rm GeV}$ or $\bar{p}_{T}^{\rm fin}>5~{\rm GeV}$, the corona
effect is manifest and the central $r_T^{\rm in}$ region is clearly
depleted w.r.t. the minimum-bias Glauber distribution (short
dashed). Imposing an additional cut on $\Delta p_T^{\rm fin}$,
$\Delta p_{T}^{\rm fin}< 0.2\,\bar{p}_{T}^{\rm fin}$, we observe the
disappearance of the corona peak together with a moderate enrichment
of the central $r_T^{\rm in}$ values and an extended ``hyper
corona'' shoulder located at $r_T^{\rm in}\approx 6-8~{\rm fm}$
where those tangential emissions take place in which both quarks
loose very little energy and therefore $\Delta p_T^{\rm fin}\approx
0$. The right hand side of  Fig. \ref{RHIC5} shows how one can
trigger on more central collisions. Increasing values of
$\bar{p}_T^{\rm fin}$ increase the sensitivity to central events.
With the condition $ \bar{p}_T^{\rm fin}>10~{\rm GeV}$, $\Delta
p_T^{\rm fin}\in [1~{\rm GeV},4~{\rm GeV}]$, the sample of events is
close to that expected from the Glauber distribution. The reason for
this is that in pQCD calculations with increasing energy of the
heavy quarks the plasma becomes more transparent, as seen on Fig.
\ref{RHIC3} (right). Energetic heavy quarks produced at small $r_T$
are hence more likely to leave the plasma with an appreciable energy
and thus they compete in number with quarks produced peripherally.
Therefore, in the right panel of Fig. \ref{RHIC5} $<r_T>$ decreases
for large $\bar{p}_T^{\rm fin}$. From Fig. \ref{RHIC5bis} (right) we
see that we nearly recover the Glauber distribution when we apply
simultaneously a $\Delta p_T^{\rm fin}<0.2\, \bar{p}_T^{\rm fin}$
and a  $\bar{p}_T^{\rm fin} > 10 GeV$  cut.
\begin{figure}[htb]
\epsfig{file=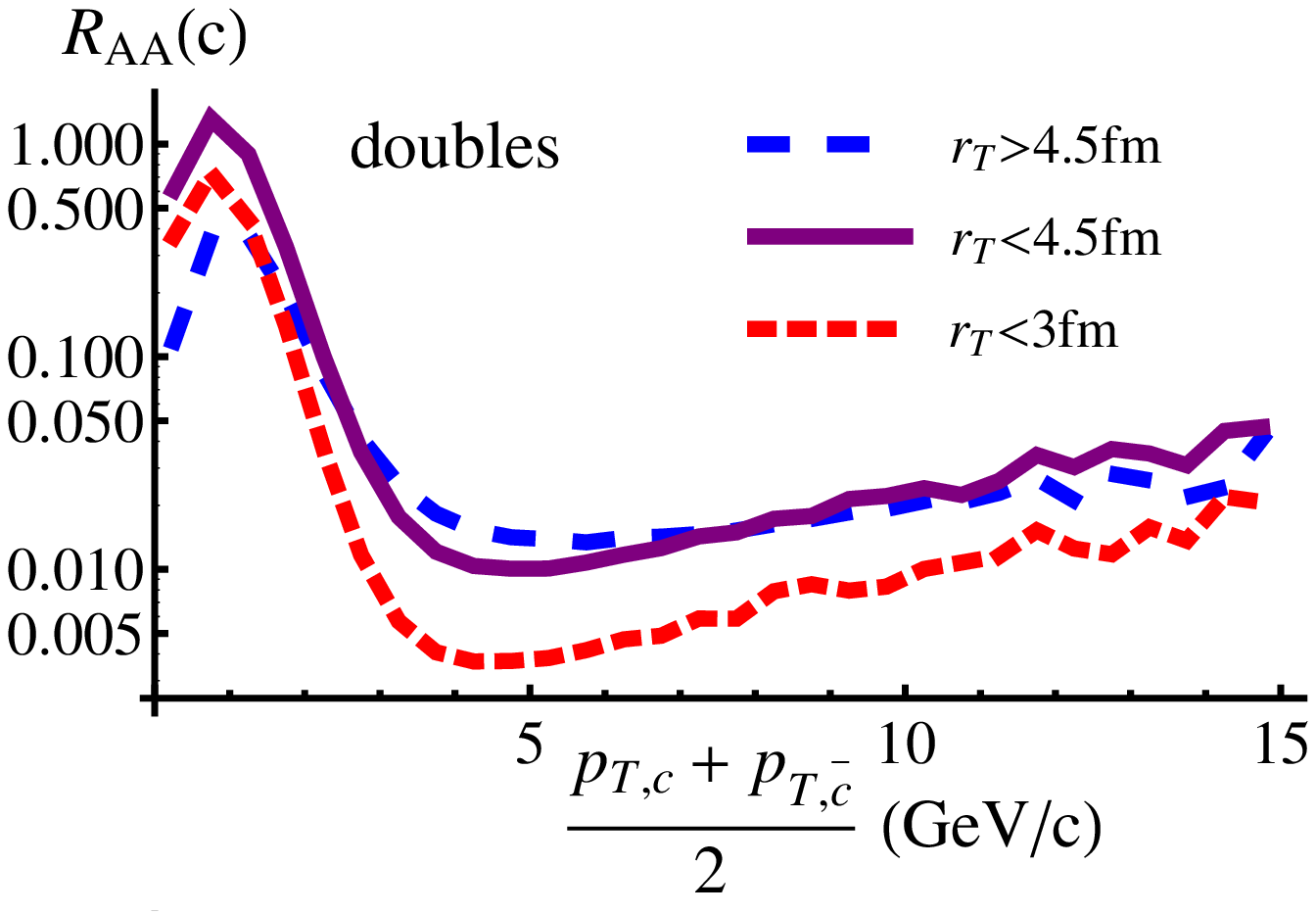,width=0.4\textwidth}
\epsfig{file=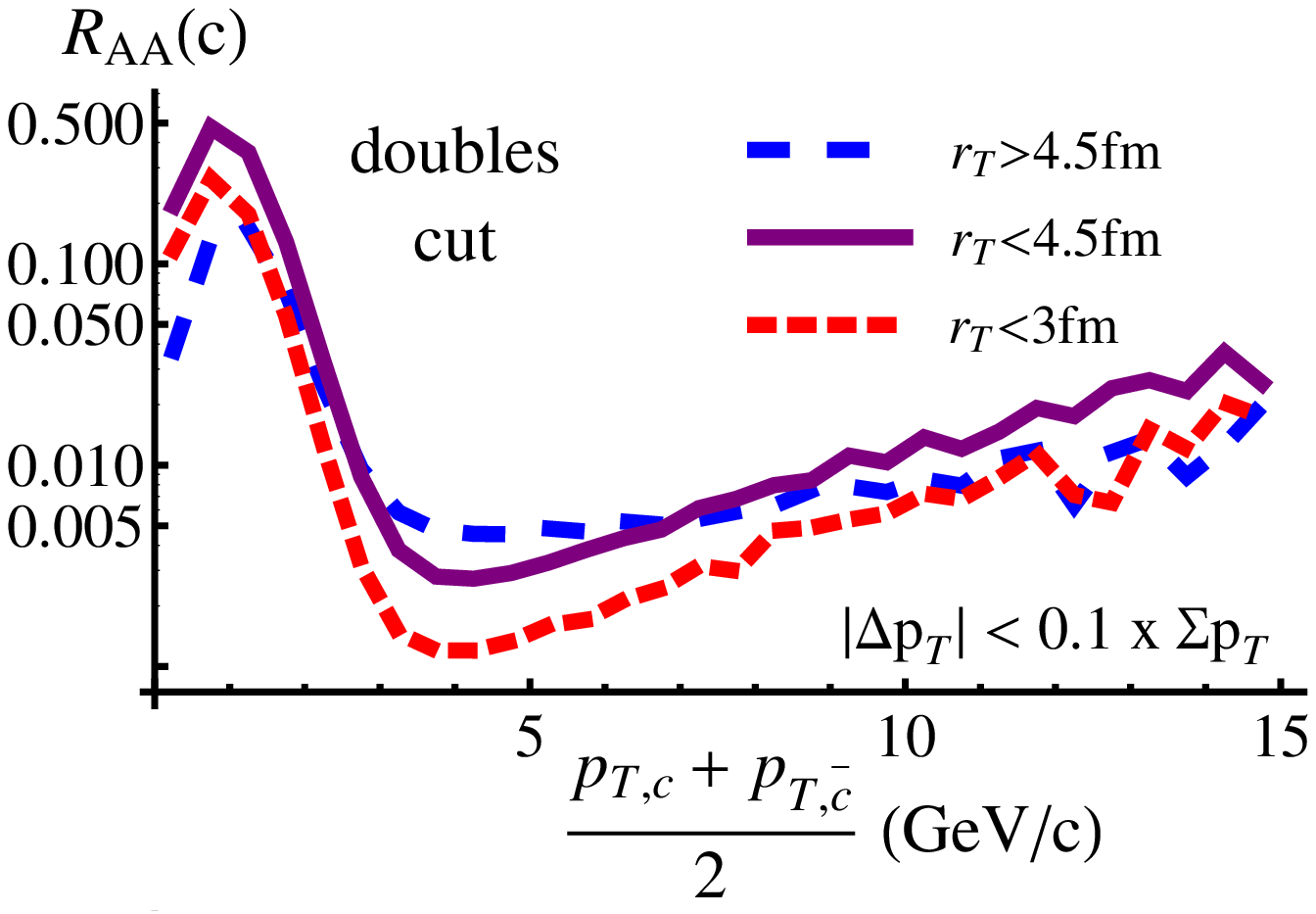,width=0.4\textwidth}
\caption{(Color online) $R_{AA}(\bar{p}_T)$ -- here defined has the
ratio $\left(dN/d\bar{p}_T^{\rm fin}\right)/
\left(dN/d\bar{p}_T^{\rm in}\right)$ as a function of the average of
the momenta of the simultaneously produced pair -- for different
conditions of the transverse momentum difference $\Delta p_T^{\rm
fin}$ between the $c$ and the $\bar{c}$ quark and various selections
on the transverse distance $r_T^{\rm in}$ of the pair creation
points with respect to the center of the reaction. The left panel
shows $R_{AA}$ for all possible $\Delta p_T^{\rm fin}$ and three
different bins of $r_T^{\rm in}$; the right panel shows the result
if we apply in addition a cut on the relative transverse momentum of
the pair.} \label{RHIC6}
\end{figure}
Another consequence of this increasing transparency is seen in Fig.
\ref{RHIC6} and Fig. \ref{RHIC6bis}. There, we display
$R_{AA}(\bar{p}_T) = \left(dN/d\bar{p}_T^{\rm fin}\right)/
\left(dN/d\bar{p}_T^{\rm in}\right)$ of $c\bar{c}$ pairs for
different $\Delta p_T^{\rm fin}$ selections in comparison with the
$R_{AA}(p_{T,c/\bar{c}})$ of single (anti)charm quarks (Fig.
\ref{RHIC6bis}). Only at small $p_T$ $dN/dp_T^{\rm in}$ differs from
$dN/d\bar{p}_T^{\rm pp}$ due to the Cronin effect, discussed in
section II.B. Therefore for larger $p_T$ $R_{AA}(\bar{p}_T)$ is a
quantity which can be measured. All curves in Fig. \ref{RHIC6} show
a minimum around $p_T \approx 5~{\rm GeV}$ where the relative energy
loss is most important. The increase beyond $p_T \approx 5~{\rm
GeV}$ becomes larger if we consider the production versus
$\bar{p}_T$ and even larger if we limit the relative transverse
momentum of the pair. While the second observation is clearly
expected, the first one, that $R_{AA}$ behaves differently as a
function of $\frac{p_{T,c}+p_{T,\bar{c}}}{2}$ than as a function of
$p_{T,c}$ is astonishing and demands some explanation in view of the
similar $\frac{dN}{d r_T}$ profiles observed in Fig. \ref{RHIC5bis}.
A detailed analysis shows that the fluctuations of the {\em average}
$\frac{p_{T,c}+p_{T,\bar{c}}}{2}$ are smaller than that of the
momentum of each of the quarks only. Therefore, the energy loss is
less washed out and this explains the $p_T$ dependence of $R_{AA}$.
\begin{figure}[H]
\begin{center}
\epsfig{file=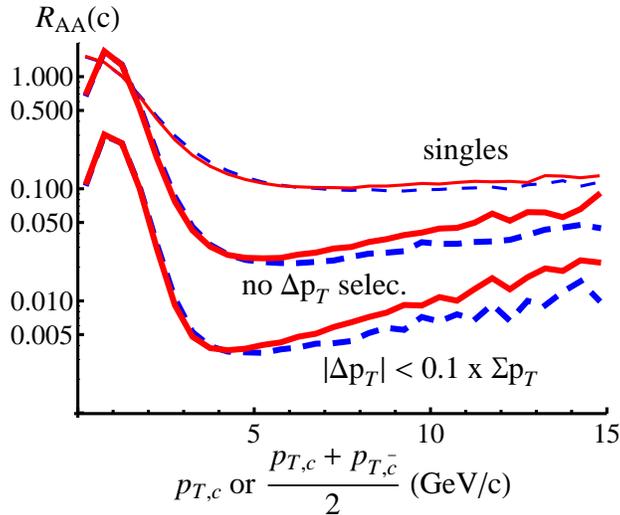,width=0.5\textwidth}
\end{center}
\caption{(Color online) $R_{AA}$ of c-quarks. Thin lines refers to
single quarks, while thick lines correspond to the $R_{AA}$ of
$\bar{p}_T$ with and without a $\Delta p_T^{\rm fin}$ selection (as
in Fig. \ref{RHIC6}). The full (dashed) line refers to model E (C)
\cite{Gossiaux:2008jv}.} \label{RHIC6bis}
\end{figure}

In conclusion, several relations can be used to validate pQCD based
models in general as well as our particular model if coincidence
data become available because the dependence of $R_{AA}$ as a
function of $\bar{p}_T$ for different $\Delta p_T^{\rm fin}$
reflects directly the interaction of the quarks with the expanding
plasma, especially the energy loss as a function of $p_T$. For
calculations with a running coupling constant (E) the increasing
transparency for large $p_T$ quarks is more pronounced than for
those with a fixed coupling constant (C). A $\Delta p_T$ cut can be
used to achieve a robust characterization of the energy loss
mechanism of heavy quarks. The heavy quark pairs are therefore one
of the few probes which are sensitive to the expansion of the plasma
and not only to its properties at the chiral/confinement phase
transition. Although we have concentrated our analysis on the heavy
{\em quarks} and not on the observed heavy mesons, we expect that
the physics seen in Fig. \ref{RHIC6bis} does not change due to
hadronization.  NLO effects at the level of $Q\bar{Q}$ production
may modify slightly the conclusions and should be included in a
future work.

We see a very complex behavior of the momentum loss of heavy quarks
in an expanding QGP. It is therefore more than questionable that
quantitative predictions of the energy loss are possible in models
which are based on the average path length of the heavy quark
trajectory in the plasma.

\subsection{Azimuthal correlations between simultaneously produced
$c$ and $\bar{c}$} Due to the large mass of the heavy quark,
interactions between a heavy quark and a plasma particle change the
direction of the heavy quark only little. We therefore expect that
\begin{figure}[H]
\begin{center}
\epsfig{file=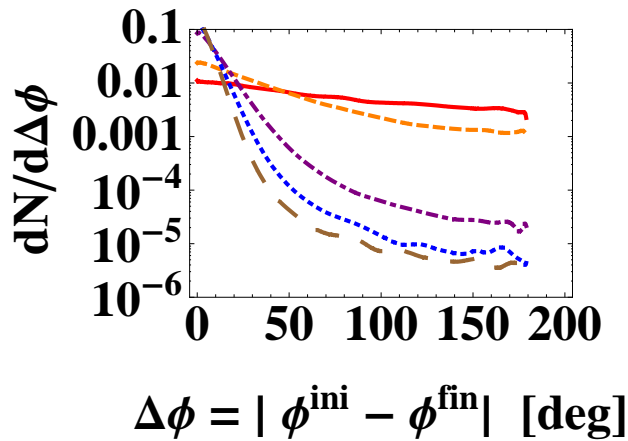,width=0.5\textwidth}
\end{center}
\caption{(Color online) Distribution of $\Delta \phi$ for different
initial $p_T$ bins [0-2 (full),2-5 (dashed),5-10
(dashed-dotted),10-15 (dotted), $>15$ (long dashed)] \ GeV.}
\label{RHIC7}
\end{figure}
the final azimuthal angle is strongly correlated with the initial
one. This is indeed the case for sufficiently high $p_T$ values, as
seen in Fig. \ref{RHIC7}. There we display the distribution of the
difference between initial and final azimuthal angle of heavy quarks
for different initial $p_T$ intervals. The higher $p_T$ the more
small angle scattering dominates and the more we see a correlation
between initial and final azimuthal angle. There is a sharp
transition towards an almost flat distribution for $p_T^{\rm
initial} < 5~{\rm GeV}$. As already seen, the kinematics allows
quarks with this initial $p_T$ to come (almost) to an equilibrium
with the environment and therefore the correlation is weakened.
\section{Comparison with AdS/CFT at RHIC}
\begin{figure}[H]
\begin{center}
\epsfig{file=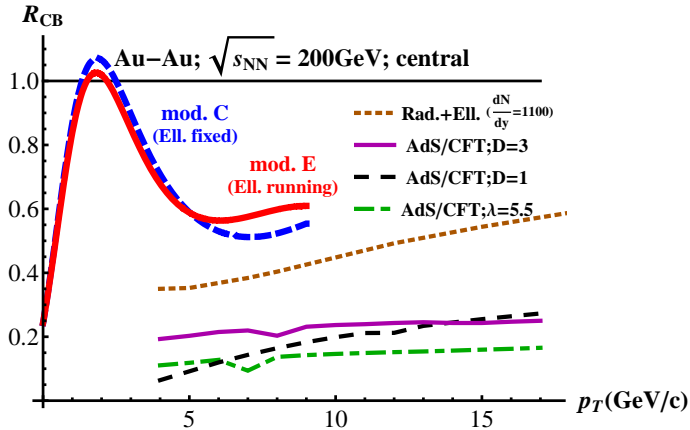,width=0.55\textwidth}
\end{center}
\caption{(Color online) $R_{CB}(p_T) = R_{AA}^c(p_T)/R_{AA}^b(p_T)$
for different theories. We compare the pQCD based ``collisional''
models C (with K=5, blue) and E (with K=1.8, red)
\cite{Gossiaux:2008jv} with the AdS/CFT calculation for different
drag coefficients $D/2\pi T$ \cite{Horowitz:2008ig} and for
$\lambda=5.5$ \cite{Horowitz:2008ig, Gubser:2006qh} as well as with
a pQCD calculation with a fixed coupling constant including
radiative collisions \cite{Wicks:2007am}. } \label{RHIC8}
\end{figure}

A completely different approach to explain the energy loss of heavy
mesons and hence of the observed low $R_{AA}$ value at high $p_T$
has recently been launched by Horowitz and Gyulassy
\cite{Horowitz:2008ig}. Their model is based on the assumption that
QCD is similar to supersymmetric Yang-Mills theory and that this
theory is dual to string theory in the limit of large $N_{color}$.
Whether this assumption is justified or not has to be verified. The
model allows to calculate the momentum loss \be \frac{dp_T}{dt}= -
const\frac{T^2}{M_q} \,p_T. \ee After having implemented this energy
loss in a Fokker-Planck approach \cite{Wicks:2007am} they could
predict quite a number of observables which can be confronted with
pQCD predictions. One of the observables for which the predictions
are quite different is the relative energy loss  of c- and b-
quarks.

pQCD calculations show a much weaker mass dependence (for a given
$p_T$). Besides a mass dependence of the energy loss in the
subdominant u-channel of the $gQ\to gQ$, which is $\propto
\frac{T^2}{m_Q^2}$, the energy loss is only mass dependent for
intermediate $p_T$ ($m_Q\ll p_T \ll \frac{m_Q^2}{T}$), where
$\frac{dE}{dx}\propto \ln(\frac{p_Q}{m_Q})$ in the the case of fixed
$\alpha_s$. Therefore the difference between the two theories can be
made evident by comparing the $R_{AA}(p_T)$ for bottom and charm
quarks. For this purpose one may define $R_{CB}(p_T) =
R_{AA}^c(p_T)/R_{AA}^b(p_T)$ \cite{Wicks:2007am}. In Fig.
\ref{RHIC8} we compared the results of the different theories. It is
evident that already the experiments at RHIC energies allow to
discard one of the theories if high $p_T$ D and B mesons could be
identified.

Our model yields quite large values of $R_{CB}$ due to the small
value of the IR regulator. pQCD calculation with a fixed coupling
constant \cite{Wicks:2007am} yield smaller values. Due to the mild
dependence on $m_Q$, mentioned before, $R_{AA}^{c}(p_T^{\rm
fin})\approx R_{AA}^{b}(p_T^{\rm fin})$ as soon as the initial
momentum distributions become similar, $\frac{dN_c}{dp_T}\approx
\frac{dN_b}{dp_T}$. This is the case for $p_T^{in} \ge 20$ GeV/c
(see Fig. \ref{pts}).

\section{Predictions for LHC}
Going from the known RHIC to the unknown LHC energy domain we are
facing the problem that it is not known how the properties of the
QGP change with increasing energy.
\begin{figure}[htb]
\epsfig{file=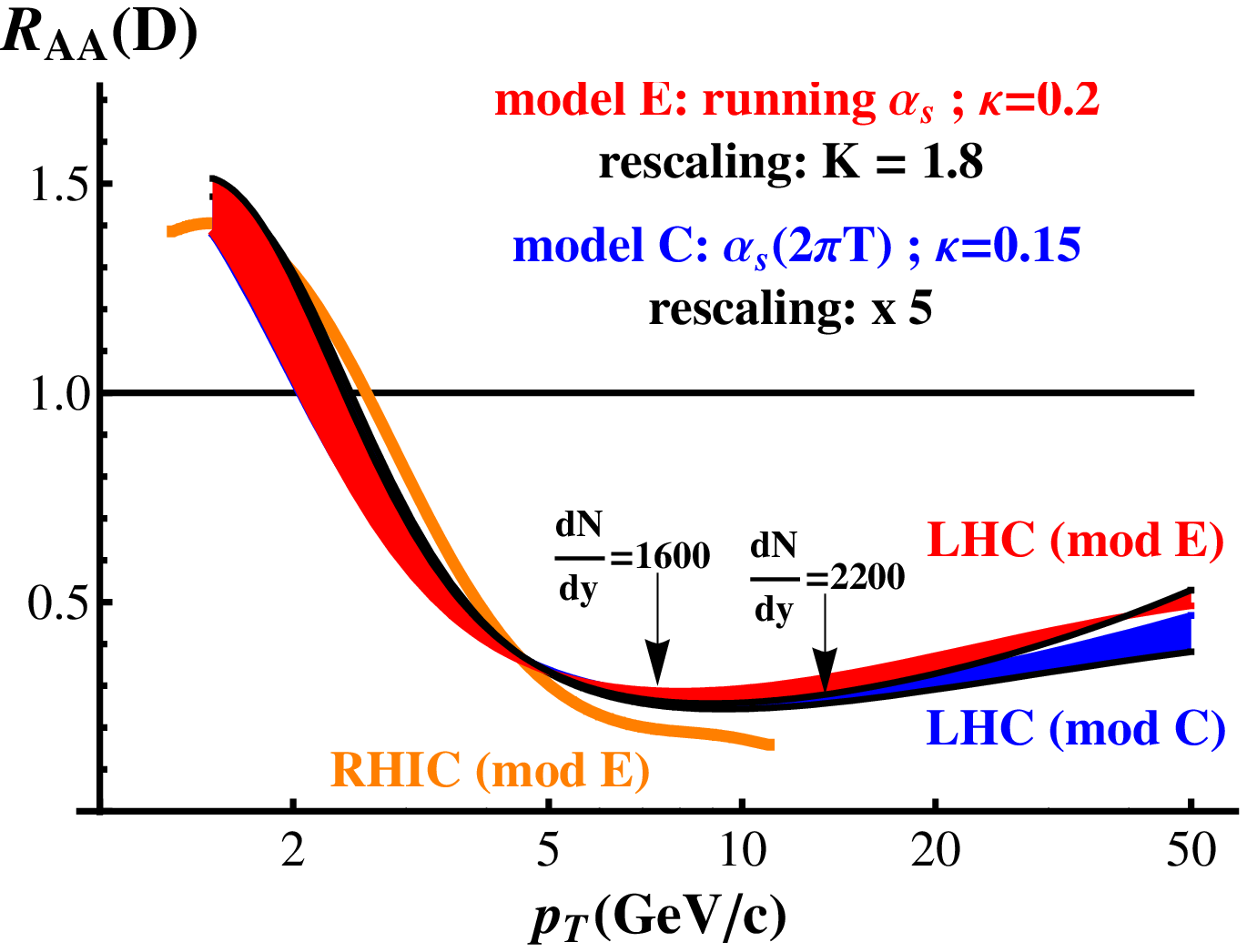,width=0.4\textwidth}
\epsfig{file=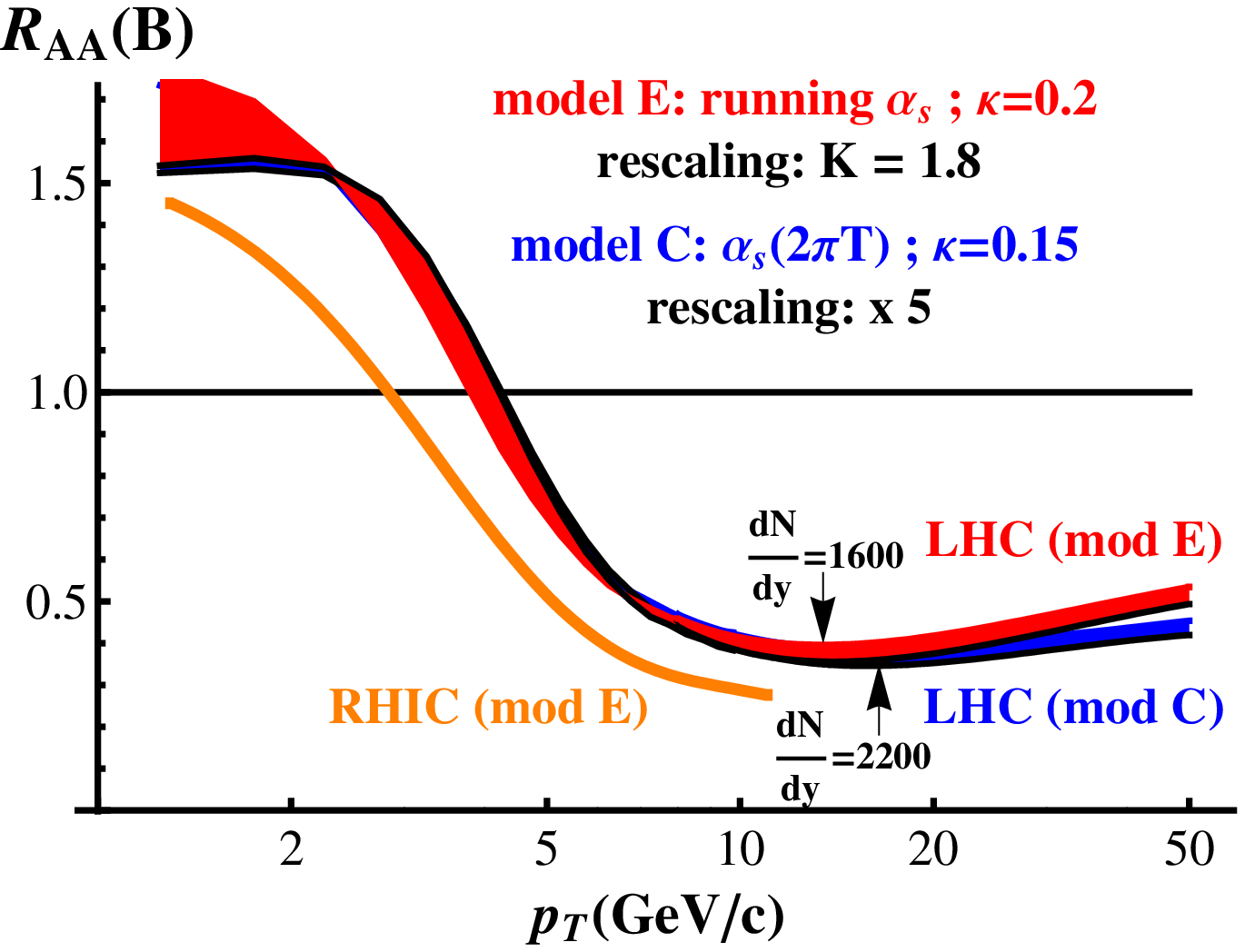,width=0.4\textwidth}
\caption{(Color online) $R_{AA}$ for central Pb+Pb collisions at 5.5
ATeV as a function of $p_T$ for D- (left) and B- (right) mesons for
model E of \cite{Gossiaux:2008jv} for LHC energies as compared to
RHIC energies.} \label{LHC1}\end{figure}

We give therefore our results for a range of charged particle
multiplicities, $1600 < dN_{\rm ch}/dy(y=0)< 2200$, which have been
predicted for LHC. We assume furthermore that the eccentricity in
coordinate space remains the same. Fig. \ref{LHC1} shows the
expected $R_{AA}$ as a function of $p_T$ for D- and B- mesons for
model E of \cite{Gossiaux:2008jv} which describes best the
experimental data at RHIC. This model has a K-factor of 1.8. We see
that at LHC we will experimentally cover the $p_T$ region in which
$R_{AA}$ increases with $p_T$. Nevertheless, the $R_{AA}$ values are
still far away from 1, expected for $p_T \to \infty$.
\begin{figure}[H]
\begin{center}
\epsfig{file=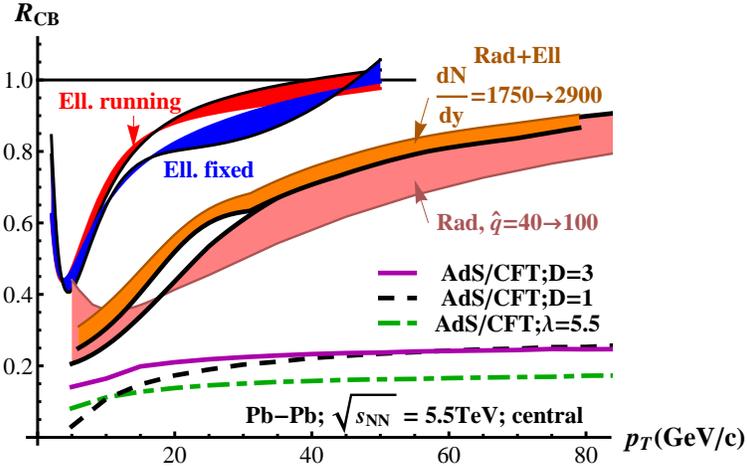,width=0.6\textwidth}
\end{center}
\caption{(Color online) $R_{CB}(p_T) = R_{AA}^c(p_T)/R_{AA}^b(p_T)$
for different theories and for central Pb+Pb collisions at 5.5 ATeV.
We compare the pQCD based models C (with K=5, blue) and E (with
K=1.8, red) \cite{Gossiaux:2008jv} with the AdS/CFT calculation for
different Drag coefficients $D/2\pi T$ \cite{Horowitz:2008ig} and
for $\lambda=5.5$ \cite{Horowitz:2008ig,Gubser:2006qh} and with the
pQCD calculation with constant coupling which includes as well
radiative energy loss \cite{Wicks:2007am}.} \label{LHC2}
\end{figure}
 The larger $p_T$ range at LHC
will make it possible to discriminate unambiguously between the
different energy loss models. Fig. \ref{LHC2} shows $R_{CB}(p_T)$
for three different theories: AdS/CFT \cite{Horowitz:2008ig}, pQCD
with radiative energy loss and constant coupling constant
\cite{Wicks:2007am} and our collisional energy loss model with a
K-factor of 1.8 (5) for running (fixed) $\alpha_s$. For moderate
$p_T$ $(p_T < 20~{\rm GeV})$ the spectral form of the c- and b-
quarks is different and  $R_{CB}(p_T)$ is far from 1, despite the
fact that the energy loss becomes more and more similar for c- and
b- quarks. Above $p_T = 30~{\rm GeV}$ an identical spectral form of
the quarks and a constant energy loss results in $R_{CB}(p_T)\approx
1$. pQCD calculation are bound to arrive finally at values of
$R_{CB}(p_T)$ close to one due to the weak mass dependence of the
cross section. The detailed form of $R_{CB}$ depends on the cross
sections or, more explicitly, on the form of the coupling constant
and of the IR regulator employed in the pQCD cross section
calculation. AdS/CFT, on the contrary, predicts even for the highest
momenta $R^{cb}=0.2-0.3$, but it has not been demonstrated yet up to
which $p_T$ values the approximations of the approach remain valid.

\begin{figure}[htb]
\epsfig{file=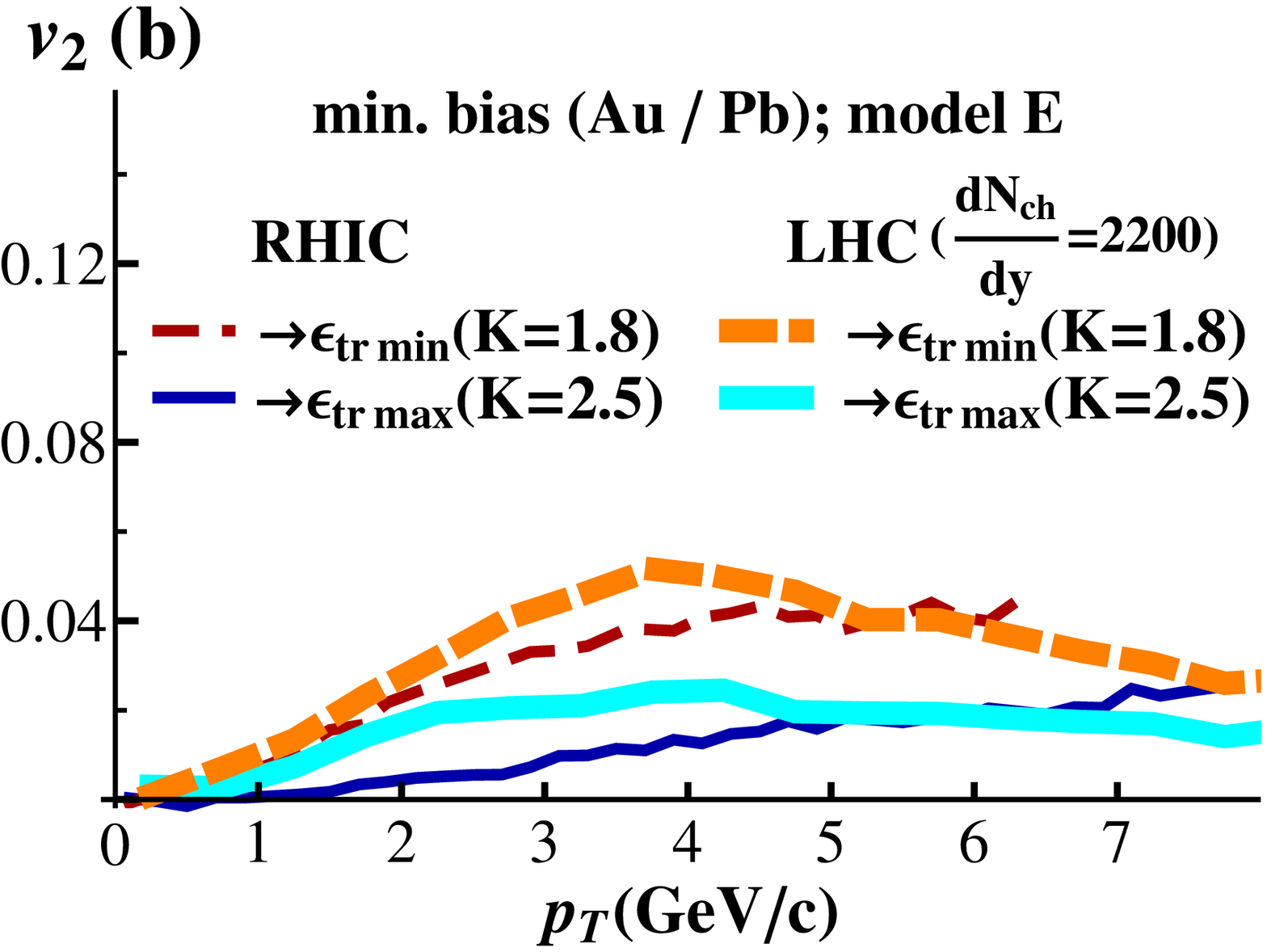,width=0.4\textwidth}
\epsfig{file=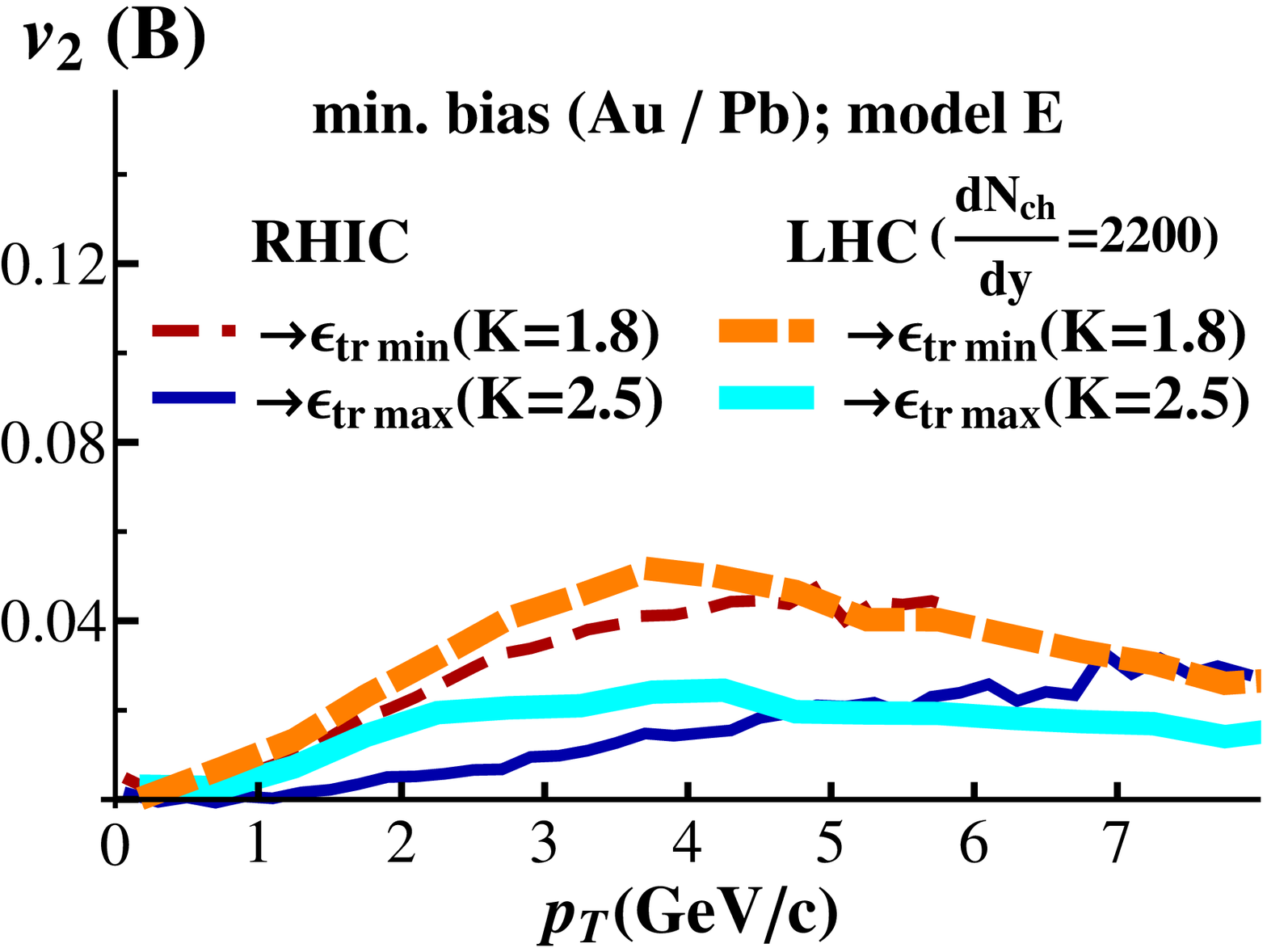,width=0.4\textwidth}
\epsfig{file=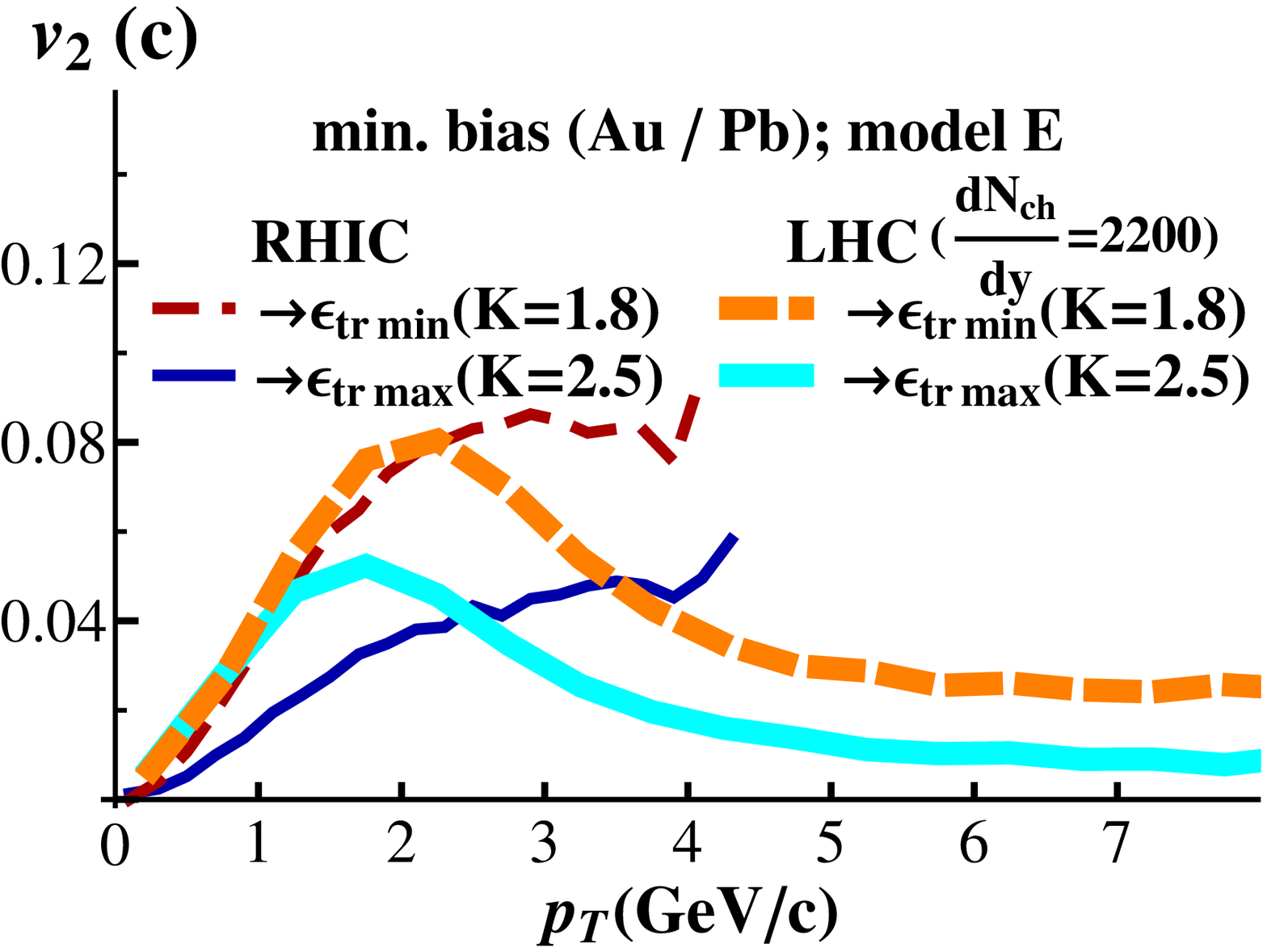,width=0.4\textwidth}
\epsfig{file=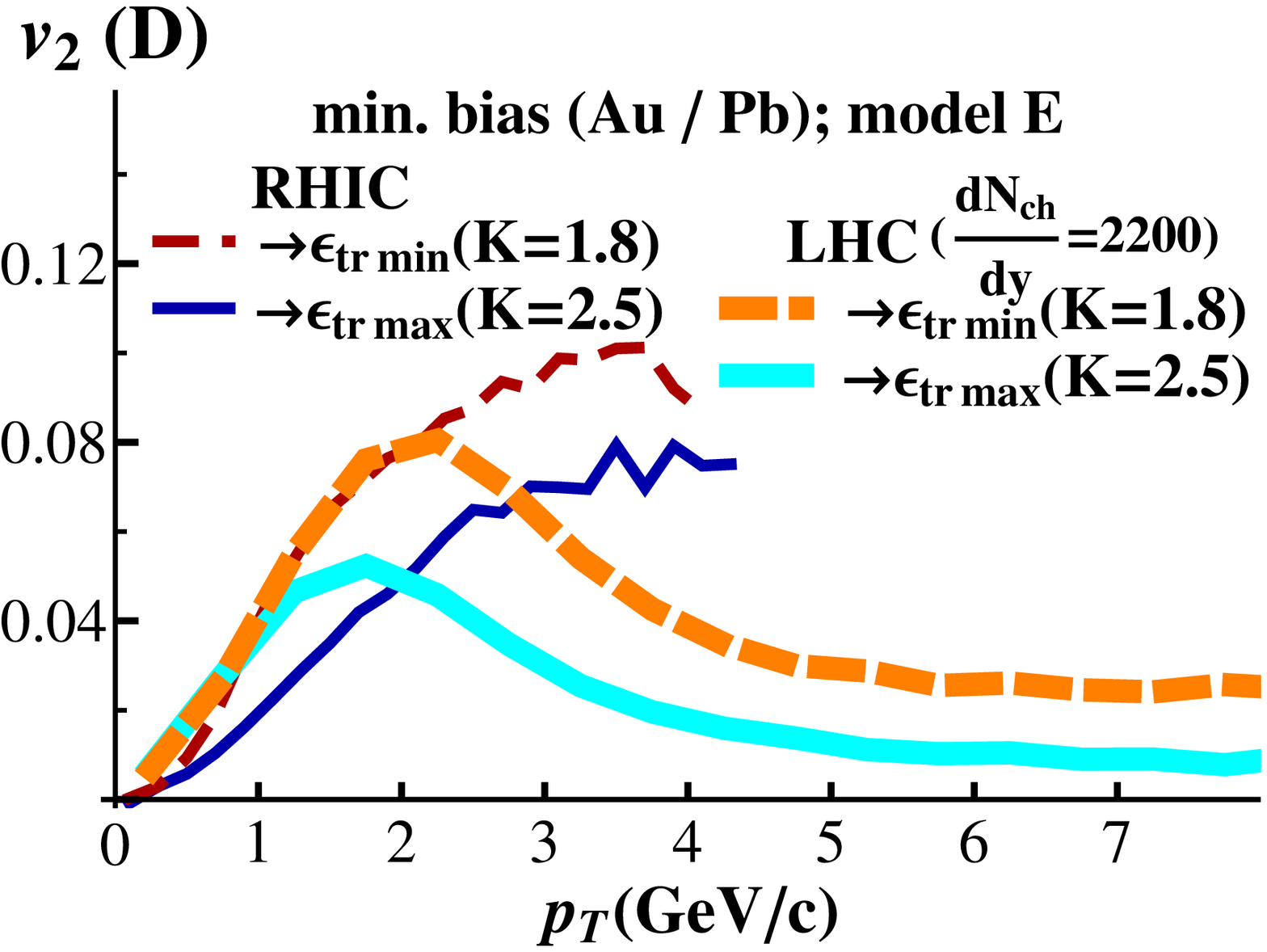,width=0.4\textwidth}
 \caption{(Color online) $v_2$ for minimum bias reactions Au+Au at 200
AGeV and Pb+Pb at 5.5 ATeV
 for model E \cite{Gossiaux:2008jv}. The K-factors are noted in the text.
 Top row: b-quarks (left) and B-mesons (right), bottom row:
c-quarks (left) and D-mesons (right).} \label{LHC3}
\end{figure}
The azimuthal anisotropy, which has been observed at RHIC energies,
will remains visible up to LHC energies, as can be inferred from
Fig. \ref{LHC3}, where we display $v_2$ for minimum bias events
separated for b-quarks (top left), B-mesons (top right), c-quarks
(bottom left) and D-Mesons (bottom right). "$\epsilon_{\rm tr \
min}$" means hadronization at the end of the mixed phase while
"$\epsilon_{\rm tr \ max}$" means hadronization at the end of the
pure QGP phase; in the latter case, a K-factor of K = 2-3, has to be
applied in order to reproduce the experimental $R_{AA}$ values for
the most central events. In \cite{Gossiaux:2008jv} we found that the
experimental values at RHIC can only be reproduced when the
hadronization takes place at the {\em end} of the mixed phase. As it
is the case for the light hadrons at RHIC, also the $v_2$ of heavy
mesons follows an hydrodynamical behavior until $p_T \approx 2~{\rm
GeV}$ but the absolute value of $v_2$ is only about half of that of
light hadrons.
\section{Conclusion}
We have described in detail the predictions of the approach which we
have advanced recently to describe the energy loss and the azimuthal
anisotropy of heavy quarks in the environment produced in heavy ion
collisions and have extended our calculation toward LHC energies. It
is based on pQCD calculations with a running coupling constant and
an IR regulator derived from hard thermal loop calculations. As shown in
ref. \cite{Gossiaux:2008jv}, with these new ingredients
the energy loss due to elastic collisions is (up to a factor of about 2)
sufficient to produce the observed $R_AA$ at RHIC collisions.
We have
presented several observables which allow to test this model.  In
particular, we predict a large azimuthal anisotropy, even at LHC
energies and strong correlations between $R_{AA}$ and the
transverse-momentum difference between the simultaneously produced
$Q\bar{Q}$ pair. Correlations between simultaneously produced heavy
quark pair will allow for triggering on central collisions. The
identification of D- and B- mesons will allow to reveal whether
AdS/CFT describes the passage of heavy quarks through matter or
whether we are still in the realm of pQCD.
\medskip

{\bf Acknowledgments:} We thank W.~Horowitz for communicating his
results and R. Vogt for communicating the $p_T$ distribution of c-
and b- quarks in pp collisions for LHC energies as well as for her
fruitful comments on a preliminary version of this article.
This work has been supported by the Agence National de la Recherche
(ANR) under the contract number ANR-08-BLAN-0093-02.


\begin{thebibliography}{11}
\bibitem{Gossiaux:2008jv}
  P.~B.~Gossiaux, J.~Aichelin
  Phys.\ Rev.\   {\bf C78}, 014904 (2008), arXiv:0802.2525 [hep-ph]

  \bibitem{Abelev:2006db}
  B.~I.~Abelev {\it et al.}  [STAR Collaboration],
  Phys.\ Rev.\ Lett.\  {\bf 98}, 192301 (2007). 


\bibitem{Adare:2006nq}
  A.~Adare {\it et al.}  [PHENIX Collaboration],
  Phys.\ Rev.\ Lett.\  {\bf 98}, 172301 (2007)  [arXiv:nucl-ex/0611018].

\bibitem{cac}
M.~Cacciari and P.~Nason,
Phys.\ Rev.\ Lett.\  {\bf 89} (2002) 122003 [arXiv:hep-ph/0204025]\\
M.~Cacciari, S.~Frixione, M.~L.~Mangano, P.~Nason and G.~Ridolfi,
JHEP {\bf 0407} (2004) 033 [arXiv:hep-ph/0312132]\\ M.~Cacciari and
P.~Nason,
JHEP {\bf 0309} (2003) 006 [arXiv:hep-ph/0306212].

\bibitem{Cacciari:2005rk}
  {}M.~Cacciari, P.~Nason and R.~Vogt
  {}Phys.\ Rev.\ Lett.\  {\bf 95}, 122001 (2005)
  [arXiv:hep-ph/0502203]
\bibitem{cacpri} R. Vogt, privat communication.

\bibitem{E866}
Jen-Chieh Peng and Mike Leitch, private communication

\bibitem{heko}
P. Kolb and U. Heinz, in Quark Gluon Plasma, World Scientific
Singapore, ed R.~Hwa and X.N.~Wang


\bibitem{Svetitsky:1987gq}
  {}B.~Svetitsky
  {}Phys.\ Rev.\  D {\bf 37}, 2484 (1988)



\bibitem{Combridge:1978kx}
  B.~L.~Combridge,
  Nucl.\ Phys.\  B {\bf 151}, 429 (1979).
\bibitem{pesh}
  A.~Peshier,
  arXiv:hep-ph/0601119 and
  {}Phys.\ Rev.\ Lett.\  {\bf 97}, 212301 (2006)
  [arXiv:hep-ph/0605294]

\bibitem{brad} E. Braaten and R. Pisarski, Phys. Rev. {\bf D45}, R1827 (1992)
\bibitem{Braaten:1991jj}
  {}E.~Braaten and M.~H.~Thoma
  {}Phys.\ Rev.\  D {\bf 44}, 1298 (1991),
  E.~Braaten and M.~H.~Thoma,
  Phys.\ Rev.\ D {\bf 44} (1991) 2625.
\bibitem{Dover:1991zn}
  {}C.~B.~Dover, U.~W.~Heinz, E.~Schnedermann and J.~Zimanyi
  {}Phys.\ Rev.\  C {\bf 44}, 1636 (1991)
\bibitem{Huang:08}
H.Z. Huang, ``Probing Properties of the QCD Medium via Heavy Quark
Induced Hadron Correlations'', talk given at the workshop
``Characterization of the Quark Gluon Plasma with Heavy Quarks'',
Bad Honnef, Germany, 26-28 June 2008,
\begin{verbatim}
http://heavy-quarks.physi.uni-heidelberg.de/Agenda/Talks/H._Huang.ppt
\end{verbatim}

\bibitem{Horowitz:2008ig}
  {}W.~A.~Horowitz and M.~Gyulassy
  {}arXiv:0804.4330 [hep-ph]

\bibitem{Gubser:2006qh}
  {}S.~S.~Gubser
  {}Phys.\ Rev.\  D {\bf 76}, 126003 (2007)
  [arXiv:hep-th/0611272]

\bibitem{Wicks:2007am}
  {}S.~Wicks, W.~Horowitz, M.~Djordjevic and M.~Gyulassy
  {}Nucl.\ Phys.\  A {\bf 783}, 493 (2007)
  [arXiv:nucl-th/0701063],
  {}S.~Wicks and M.~Gyulassy
  {}arXiv:nucl-th/0701088
\end{thebibliography}
\end{document}